\documentclass[prd,amsmath,groupedaddress,onecolumn,eqsecnum,nofootinbib]{revtex4-2}
\usepackage{tikz}
\usepackage{natbib}
\usepackage{hyperref}
\usepackage{subfig}
\usepackage{graphicx}
\usepackage{lipsum}
\usepackage{enumerate}
\usepackage{lipsum, babel}
\usepackage{graphicx}
\usepackage{dcolumn}
\usepackage{amsmath}
\usepackage{amssymb}
\usepackage{amsfonts}
\usepackage{orcidlink}
\usepackage{bm}
\usepackage{float}
\hypersetup{colorlinks=true, linkcolor=blue, citecolor=blue, urlcolor=blue}
\usepackage{caption}
\usepackage[hmargin={0.35in,0.35in},height=9.5in, top=0.7in
]{geometry}

\begin{document}

\title{Photon Deflection and Magnification in Kalb-Ramond Black Holes with Topological String Configurations}

\author{Faizuddin Ahmed\orcidlink{0000-0003-2196-9622}}
\email{faizuddinahmed15@gmail.com (Corresp. author)}
\affiliation{Department of Physics, University of Science and Technology Meghalaya, Ri-Bhoi, 793101, Meghalaya,
India
}

\author{\.{I}zzet Sakall{\i}\orcidlink{0000-0001-7827-9476}}
\email{izzet.sakalli@emu.edu.tr}
\affiliation{Department of Physics, Eastern Mediterranean University, Famagusta Northern Cyprus, 99628,  via Mersin 10, Turkiye
}
\author{Ahmad Al-Badawi\orcidlink{0000-0002-3127-3453}}
\email{ahmadbadawi@ahu.edu.jo}
\affiliation{Department of Physics, Al-Hussein Bin Talal University, Ma'an, 71111, Jordan
}

\date{\today}

\begin{abstract}
This theoretical investigation examines gravitational lensing phenomena in Schwarzschild-like black holes (BHs) within Kalb-Ramond (KR) gravity frameworks incorporating cosmic string (CS) and cloud of strings (CoS) topological configurations. We develop comprehensive analytical methodologies for investigating photon deflection angles and magnification characteristics in both Schwarzschild-like BHs in KR gravity pierced by CSs (SKRCS) and Schwarzschild-like BHs in KR gravity with CoS (SKRCoS) spacetime geometries through dual approaches: perturbative expansions yielding approximate solutions and exact elliptic integral formulations providing complete mathematical descriptions across parameter spaces. For CS configurations characterized by Lorentz violation (LV) and CS parameters, deflection angles exhibit systematic modifications through composite geometric factors, while CoS geometries demonstrate distinct deflection characteristics incorporating additional topological parameters that fundamentally alter light propagation dynamics. Magnification analysis reveals distinctive critical curve positioning modifications and amplitude scaling relationships enabling observational discrimination between exotic BH scenarios and conventional spacetime geometries. Strong field analysis utilizing established mathematical frameworks establishes logarithmic divergence coefficients characterizing fundamental scaling behavior in photon sphere proximity regimes. Observational constraints from Solar System precision tests restrict LV parameters within stringent bounds, while galactic-scale observations permit expanded parameter ranges for CS and CoS configurations. 
\end{abstract}



\maketitle

{\small

\section{Introduction}\label{isec1}

The theoretical exploration of modified gravitational theories represents one of the most profound frontiers in contemporary theoretical physics, wherein classical general relativity (GR) undergoes systematic extensions to incorporate exotic physical mechanisms that potentially address fundamental questions in cosmology and quantum gravity \cite{is01,is02,is03}. Among the most compelling theoretical frameworks for investigating such modifications are spacetime geometries that systematically incorporate LV mechanisms through KR field configurations, establishing novel gravitational paradigms that transcend conventional Einstein-Hilbert formulations while maintaining mathematical precision and observational consistency \cite{is04,is05,is06}.

The KR field framework, originally formulated within string theory contexts, introduces a fundamental antisymmetric tensor field $\mathcal{B}^{\mu\nu}$ that couples nonminimally to gravitational degrees of freedom, inducing spontaneous LV through dynamical symmetry breaking mechanisms \cite{is07,is08}. This theoretical foundation establishes Schwarzschild-like BH solutions that systematically deviate from conventional spacetime geometries through the incorporation of dimensionless LV parameter $\ell$, characterizing the magnitude of fundamental symmetry violations arising from non-trivial VEV configurations of the underlying KR field \cite{is09,is10,is10x,is10xx}. The resulting modified spacetime geometries exhibit profound alterations in gravitational lensing characteristics, offering unprecedented theoretical pathways for investigating exotic physics through precision astronomical observations of photon deflection phenomena.

The systematic incorporation of topological defect structures into KR gravity frameworks represents a significant theoretical advancement, establishing novel BH configurations that simultaneously incorporate both fundamental symmetry violations and cosmologically motivated string theory constructs \cite{is11,is12,is13}. CS configurations, characterized by one-dimensional topological defects with characteristic tension parameter $\mu$, introduce additional geometric modifications through the CS parameter $\beta = (1-4G\mu)$, systematically altering spacetime curvature in the vicinity of gravitational sources \cite{is14,is15}. The theoretical synthesis of KR modifications with CS topological structures yields SKRCS BH configurations that exhibit distinctive gravitational lensing signatures, potentially distinguishable from conventional Schwarzschild predictions through systematic analysis of deflection angle modifications and magnification characteristics.

Furthermore, the theoretical framework extends to encompass CoS distributions, representing alternative topological configurations where extended string matter sources permeate the spacetime geometry through distributed energy-momentum contributions \cite{is16,is17}. The SKRCoS BH configurations incorporate CoS effects through the parameter $\alpha$, establishing composite modifications $\eta = \frac{1}{1-\ell} - \alpha$ that systematically encode both LV mechanisms and distributed string matter sources within unified theoretical paradigms \cite{is18,is19}. The theoretical investigation of gravitational lensing in modified spacetime geometries necessitates particular analytical methodologies that systematically account for both weak and strong field regimes, wherein photon trajectories exhibit fundamentally different scaling behaviors and mathematical characteristics \cite{is20,is21,is22}. Weak field deflection angle analysis employs perturbative expansion techniques to obtain approximate analytical expressions valid for impact parameters significantly larger than the gravitational radius, establishing systematic corrections to classical Schwarzschild predictions that encode the influence of exotic physics parameters \cite{is23,is24}. Conversely, strong field deflection analysis requires sophisticated treatment of logarithmic divergences that emerge as photon trajectories approach the photon sphere, necessitating some mathematical techniques such as the Bozza formalism to extract meaningful physical predictions from otherwise singular behavior \cite{is25,is26}.

The magnification analysis of gravitational lensing phenomena provides complementary theoretical insights into the nature of modified spacetime curvature, wherein the cross-sectional area modifications of photon ray bundles directly reflect the underlying geometric deformations induced by exotic physics contributions \cite{is27,is28}. The systematic decomposition of magnification effects into tangential and radial components reveals distinctive signatures of LV and topological defect modifications, establishing critical curve positioning and amplitude scaling relationships that potentially enable observational discrimination between different theoretical paradigms \cite{is29,is30,is30x,is30xx,is30xxx}. These magnification characteristics exhibit systematic parameter dependencies that could provide quantitative frameworks for constraining exotic physics through statistical analysis of large gravitational lensing surveys.

The primary motivation for this investigation emerges from the fundamental recognition that precision gravitational lensing observations potentially offer unprecedented sensitivity to exotic physics modifications that remain undetectable through conventional gravitational tests \cite{is31,is32,is33}. Contemporary astronomical surveys achieve extraordinary precision in astrometric measurements, establishing observational capabilities that approach the theoretical sensitivity required for detecting subtle deviations from GR predictions in both weak and strong field regimes \cite{is34,is35}. The systematic theoretical characterization of gravitational lensing in SKR BH configurations therefore establishes essential foundations for optimizing observational strategies aimed at constraining fundamental physics beyond standard theoretical frameworks.

The specific research objectives encompass comprehensive theoretical analysis of deflection angle modifications in both SKRCS and SKRCoS configurations, employing dual analytical methodologies that provide both approximate perturbative solutions and exact elliptic integral representations \cite{is36,is37}. The investigation explores parameter space dependencies to establish quantitative relationships between observable lensing characteristics and underlying exotic physics parameters, providing theoretical frameworks for experimental constraint of LV mechanisms and topological defect contributions \cite{is38,is39}. Furthermore, the research establishes magnification analysis across complete parameter spaces for both SKRCS and SKRCoS configurations, revealing some modifications to critical curve positioning and amplitude scaling that potentially enable observational discrimination between exotic BH scenarios and conventional spacetime geometries \cite{is40,is41}. The strong field deflection analysis employs especial techniques to extract logarithmic divergence coefficients that characterize the fundamental scaling behavior in regimes where conventional approximations fail, establishing quantitative predictions for precision gravitational lensing observations around the exotic BH configurations.

This paper is organized as follows: Section~\ref{isec2} establishes the theoretical foundations of SKRCS BH geometry and develops comprehensive deflection angle analysis through both perturbative and exact analytical methodologies. Section~\ref{isec3} presents systematic magnification analysis for SKRCS configurations, exploring parameter dependencies across observationally constrained and theoretically motivated regimes. Section~\ref{isec4} extends the theoretical framework to encompass SKRCoS BH configurations, developing parallel analytical treatments for deflection angles in modified spacetime geometries incorporating CoS distributions. Section~\ref{isec5} provides comprehensive magnification analysis for SKRCoS scenarios, establishing comparative frameworks for understanding the relative influence of different exotic physics contributions. Section~\ref{isec6} investigates strong field deflection phenomena using advanced mathematical techniques to characterize logarithmic divergence behavior in both SKRCS and SKRCoS configurations. Finally, Section~\ref{isec7} synthesizes the theoretical findings and discusses implications for future observational programs aimed at constraining exotic gravitational physics through precision gravitational lensing measurements.

\section{SKRCS BH Geometry and Deflection of light}\label{isec2}

The theoretical framework of KR gravity represents a fundamental departure from conventional GR through the implementation of spontaneous LV mechanisms. The KR solution, originally formulated in \cite{is09}, establishes a Schwarzschild-type spacetime geometry that incorporates systematic breaking of Lorentz symmetry. This modified gravitational framework emerges from the non-zero VEV of the KR field permeating spacetime, providing a solid theoretical foundation for investigating exotic gravitational phenomena \cite{is10,is07}.

The KR BH solution is characterized by the following line element:
\begin{equation}
    ds^2=-f(r,\ell)\,dt^2+\frac{dr^2}{f(r,\ell)}+r^2\,(d\theta^2+\sin^2 \theta\,d\phi^2),\label{a1}
\end{equation}
where the metric function incorporates LV modifications:
\begin{equation}
    f(r,\ell)=\frac{1}{1-\ell}-\frac{2\,M}{r}.\label{a2}
\end{equation}

The dimensionless parameter $\ell$ quantifies the magnitude of Lorentz symmetry violation arising from the non-trivial KR field configuration. Recent gravitational lensing investigations of this BH solution have revealed distinctive observational signatures that potentially distinguish KR gravity from conventional theoretical frameworks \cite{is08}.

Observational constraints from Solar System precision tests impose stringent bounds on the LV parameter through multiple independent measurements. The Shapiro time delay experiments yield the most restrictive constraints: $-6.1 \times 10^{-13} \leq \ell \leq 2.8 \times 10^{-14}$ \cite{is06}. Complementary constraints from Mercury's perihelion precession and light deflection measurements provide additional verification of these bounds.

\begin{table}[h!]
\centering
\resizebox{0.5\textwidth}{!}{%
\def\arraystretch{1.4}
\begin{tabular}{ll}
\hline
\textbf{Solar tests} & \textbf{Constraints} \\
\hline
Mercury precession & $-3.7 \times 10^{-12} \leq \ell \leq 1.9 \times 10^{-11}$ \\
Light deflection & $-1.1 \times 10^{-10} \leq \ell \leq 5.4 \times 10^{-10}$ \\
Shapiro time-delay & $-6.1 \times 10^{-13} \leq \ell \leq 2.8 \times 10^{-14}$ \\
\hline
\end{tabular}
}
\caption{Constraints on the LV parameter $\ell$ from Solar System tests \cite{is09}.}
\end{table}

However, galactic-scale observations provide fundamentally different constraint regimes. Analysis of the Sgr A* supermassive BH at the Galactic center yields significantly relaxed bounds: $-0.18502 < \ell < 0.06093$ \cite{is32}. This dramatic scale-dependent variation in constraint tightness suggests potential theoretical pathways for detecting LV effects in astrophysical environments where conventional Solar System bounds may not apply. Furthermore, extragalactic observations of supermassive BHs in environments characterized by extreme gravitational fields and enhanced matter-energy densities may permit even broader parameter exploration regimes, potentially enabling detection of LV signatures that remain unobservable within local galactic constraints and establishing novel observational frameworks for probing fundamental spacetime modifications in cosmologically distant astrophysical systems.

\begin{table}[h]
\centering
\resizebox{0.5\textwidth}{!}{%
\def\arraystretch{1.4}
\begin{tabular}{cccc}
\hline \hline
\multicolumn{4}{c}{Parameter values}                                                                               \\ \hline
\hspace{1 mm}Survey\hspace{1 mm} & \hspace{1 mm}$M (\times 10^6 M_{\odot})$\hspace{1 mm} & \hspace{1 mm}$D$\hspace{1 mm} (kpc) & \hspace{1 mm}Reference\hspace{1 mm} \\ 
Keck                             & $3.951 \pm 0.047$                                     & $7.953 \pm 0.050 \pm 0.032$         & \cite{is33} \\ \hline
\end{tabular}%
}
\caption{Sgr A* mass and distance as inferred by the Keck and VLTI observations.}
\label{tab:tabela1}
\end{table}

The integration of topological defects into KR gravity frameworks represents a significant theoretical advancement. CS-pierced Schwarzschild spacetimes, originally investigated in \cite{is14,aa2}, introduce additional geometric modifications through the CS parameter $\beta$. The CS-modified line element takes the form:
\begin{equation}
    ds^2=-f(r)\,dt^2+\frac{dr^2}{f(r)}+r^2\,(d\theta^2+\beta^2\,\sin^2 \theta\,d\phi^2),\label{a3}
\end{equation}
with the standard Schwarzschild metric function:
\begin{equation}
    f(r)=1-\frac{2\,m}{r},\label{a4}
\end{equation}
where the physical BH mass is given by $M=\beta\,m$, and $\beta=(1-4\,G\,\mu)$ represents the CS parameter with $\mu$ denoting the mass per unit length of the string configuration.

The theoretical synthesis of KR gravity modifications with CS geometries yields the SKRCS BH configuration, representing a novel class of modified spacetimes with potentially observable gravitational lensing signatures. The SKRCS line element incorporates both LV and topological defect contributions:
\begin{eqnarray}
   ds^2=-\mathcal{A}(r,\ell)\,dt^2+\frac{dr^2}{\mathcal{A}(r,\ell)}+r^2\,(d\theta^2+\beta^2\,\sin^2 \theta\,d\phi^2),\label{b1}
\end{eqnarray}
where the modified metric function becomes:
\begin{equation}
    \mathcal{A}(r,\ell)=\delta-\frac{2\,M}{\beta r},\label{b2}
\end{equation}
with $\delta=\frac{1}{1-\ell}$, $M=\beta\,M_0$ representing the physical BH mass, and $M_0$ denoting the bare Schwarzschild mass parameter.

This theoretical framework exhibits elegant limiting behavior: $\beta=1$ recovers the pure KR BH solution \cite{is09}, while $\ell=0$ yields the standard CS-pierced Schwarzschild geometry \cite{is14}. The simultaneous limits $\ell=0$ and $\beta=1$ reproduce conventional Schwarzschild spacetime.

The investigation of photon trajectories in SKRCS geometry requires systematic analysis of null geodesics through variational principles. The Lagrangian density function for particle motion is defined as:
\begin{equation}
\mathcal{L}=\frac{1}{2}\,g_{\mu\nu}\,\left(\frac{dx^{\mu}}{d\tau}\right)\,\left(\frac{dx^{\nu}}{d\tau}\right),\label{b3}
\end{equation}
where $\tau$ represents the affine parameter and $g_{\mu\nu}$ denotes the metric tensor components.

Substituting the SKRCS metric (\ref{b1}), the explicit Lagrangian becomes:
\begin{equation}
\mathcal{L}=\frac{1}{2}\,\Bigg[-\mathcal{A}\,\left(\frac{dt}{d\tau}\right)^2+\frac{1}{\mathcal{A}}\,\left(\frac{dr}{d\tau}\right)^2+r^2\,\left(\frac{d\theta}{d\tau}\right)^2+\beta^2\,r^2\,\sin^2\theta\,\left(\frac{d\phi}{d\tau}\right)^2\Bigg].\label{b4}
\end{equation}

Application of the Euler-Lagrange equations yields the fundamental geodesic relations:
\begin{eqnarray}
    &&\frac{dt}{d\tau}=\frac{\mathrm{E}}{\mathcal{A}}\,,\nonumber\\
    &&\frac{d\phi}{d\tau}=\frac{\mathrm{L}_z}{\beta\,r^2\,\sin^2 \theta},\nonumber\\
    &&\frac{d\theta}{d\tau}=\frac{1}{r^2}\,\sqrt{\mathrm{L}^2-\frac{\mathrm{L}^2_z}{\sin^2 \theta}}.\label{b5}
\end{eqnarray}

Here, $\mathrm{E}$ represents the conserved energy parameter, $\mathrm{L}_z=p_{\phi}/\beta$ denotes the modified angular momentum, and $\mathrm{L}^2=p^2_{\theta}+p^2_{\phi}\,(\beta\,\sin \theta)^{-2}$ characterizes the total angular momentum magnitude.

For null geodesics, the constraint equation becomes:
\begin{equation}
    \left(\frac{dr}{d\tau}\right)^2+V_\text{eff}(r)=\mathrm{E}^2,\label{b6}
\end{equation}
where the effective potential incorporates both LV and CS modifications:
\begin{equation}
    V_\text{eff}(r)=\frac{\mathrm{L}^2}{r^2}\,\mathcal{A}=\frac{\mathrm{L}^2}{r^2}\,\left(\frac{1}{1-\ell}-\frac{2\,M}{\beta\,r}\right).\label{b7}
\end{equation}

The angular coordinate evolution exhibits complex parameter dependencies through:
\begin{equation}
    \frac{d\theta}{d\phi}=\beta\,\sin \theta\,(\zeta^2\,\sin^2 \theta-1)^{1/2},\label{b8}
\end{equation}
where $\zeta^2=\mathrm{L}^2/\mathrm{L}^2_z$ represents a dimensionless parameter governing orbital characteristics.

Simplification yields the fundamental angular relationship:
\begin{equation}
    \cot^2 \theta=(\zeta^2-1)\,\sin^2 (\beta\,\phi).\label{b9}
\end{equation}

This expression demonstrates the intricate coupling between the angular coordinates $\theta$ and $\phi$, the CS parameter $\beta$, and the orbital parameter $\zeta$. The parameter $\zeta$ fundamentally distinguishes geodesic motions from pure equatorial trajectories.

For equatorial motion ($\theta=\pi/2$), the condition $\zeta=1 \Rightarrow \mathrm{L}_z=\mathrm{L}$ yields $\cot \theta=0$, confirming confinement to the equatorial plane. In this configuration, the radial orbit equation becomes:
\begin{equation}
    \left(\frac{dr}{d\phi}\right)^2=\beta^2\,r^4\,\left[\frac{1}{\gamma^2}-\frac{1}{r^2}\,\left(\frac{1}{1-\ell}-\frac{2\,M}{\beta\,r} \right)\right],\label{b10}
\end{equation}
where $\gamma=\mathrm{L}/\mathrm{E}$ represents the impact parameter characterizing photon trajectory geometry.

This foundational framework establishes the theoretical basis for subsequent analysis of deflection angles through both perturbative and exact analytical methodologies, providing comprehensive tools for investigating gravitational lensing phenomena in SKRCS spacetimes.

\subsection{Deflection Angle of SKRCS BH via Perturbation Method} \label{subsec2.1}

The theoretical determination of photon deflection angles in SKRCS geometries requires sophisticated analytical methodologies that systematically account for both LV and CS modifications to conventional gravitational lensing. The perturbative approach provides a powerful framework for obtaining approximate solutions with controlled accuracy, enabling systematic exploration of parameter-dependent modifications to standard GR predictions \cite{sec2is10}.

The analysis begins with the implementation of the canonical variable transformation $r(\phi)=\frac{1}{u(\phi)}$ applied to the fundamental orbit equation (\ref{b10}). This transformation systematically converts the radial coordinate dependence into an inverse radial distance formulation, yielding enhanced mathematical tractability for subsequent perturbative analysis \cite{is23,Sucu:2025lqa}.

Through systematic algebraic manipulation, the orbit equation transforms into a second-order differential equation governing the inverse radial distance $u$ from the SKRCS BH:
\begin{equation}
    \frac{d^2u}{d\phi^2}+\beta^2 \delta u=3M\beta u^2, \label{izpp1}
\end{equation}

The mathematical structure of Eq. \eqref{izpp1} explicitly reveals the fundamental mechanism through which the LV parameter $\ell$ introduces systematic perturbations to the classical Schwarzschild lensing framework. The modified frequency term $\beta^2 \delta$ incorporates both CS and LV contributions, distinguishing this formulation from conventional gravitational lensing theories \cite{is25}.

The analytical solution of Eq. \eqref{izpp1} employs a comprehensive perturbative methodology based on power series expansion in the gravitational coupling strength. The perturbative ansatz takes the form:
\begin{equation}  \label{izp}
u = u_0 + \varepsilon u_1 + \varepsilon^2 u_2 +\cdots,
\end{equation}

where $\varepsilon$ represents the dimensionless gravitational strength parameter, physically corresponding to the ratio of the gravitational radius to the characteristic length scale of the photon trajectory. This expansion methodology enables systematic approximation to arbitrary order while maintaining analytical tractability \cite{is24}.

The leading-order or homogeneous equation characterizes the unperturbed photon trajectory in the modified spacetime geometry:
\begin{equation}
\frac{d^2u_0}{d \phi^2}+\beta^2 \delta u_0=0. \label{izm1}
\end{equation}

The analytical solution of this harmonic oscillator equation yields:
\begin{equation}
u_0 = u_\bot\cos (\beta \sqrt{\delta} \phi),
\end{equation}

where $u_\bot$ represents the inverse of the Newtonian closest approach distance, with the identification $u_\bot = 1/\gamma$ relating to the impact parameter at asymptotic infinity. This zeroth-order solution establishes the fundamental trajectory characteristics in the absence of higher-order gravitational corrections.

The systematic derivation of higher-order corrections requires the solution of inhomogeneous differential equations incorporating the nonlinear gravitational source terms. The first-order and second-order correction equations are:
\begin{align}
&\frac{d^2u_1}{d \phi^2} + \beta^2 \delta u_1 = 3 u_\bot M \beta \cos^2 (\beta \sqrt{\delta} \phi), \label{iz6} \\
&\frac{d^2u_2}{d \phi^2} +  \beta^2 \delta u_2 =6 M \beta \cos (\beta \sqrt{\delta} \phi) u_1. \label{iz7}
\end{align}

These inhomogeneous differential equations admit closed-form analytical solutions through standard Green's function techniques. The first-order correction incorporates the leading gravitational modification:
\begin{equation} \label{eq:u1_sol}
    u_1 = \frac{1}{2} \frac{u_\bot \left(3-\cos (2 \beta \sqrt{\delta} \phi)\right)}{\delta \beta},
\end{equation}

while the second-order correction captures more subtle gravitational effects:
\begin{align} \label{eq:u2_sol}
    u_2 &= \frac{3}{16} \frac{u_\bot\left(\cos (3 \beta \sqrt{\delta} \phi) \sqrt{\delta}+20 \sin (\beta \sqrt{\delta}  \phi) \phi \beta \delta\right)}{\delta^{(5 / 2)} \beta^2}.
\end{align}

The systematic combination of perturbative orders according to Eq. \eqref{izp} yields the comprehensive trajectory solution:
\begin{align}
u =\ & u_\bot \left[\cos(\beta \sqrt{\delta} \phi) 
+\frac{\varepsilon \left(3 - \cos(2 \beta \sqrt{\delta} \phi)\right)}{2\delta \beta} 
 + \frac{3\varepsilon^2 \left( \cos(3 \beta \sqrt{\delta} \phi) \sqrt{\delta} + 20 \sin(\beta \sqrt{\delta} \phi) \phi \beta \delta \right)}{16\delta^{5/2} \beta^2}\right].
\end{align}

This expression systematically incorporates LV and CS modifications at multiple perturbative orders, providing a comprehensive framework for analyzing photon trajectories in SKRCS geometries with controlled approximation accuracy.

The calculation of the asymptotic deflection angle $\alpha_{\circledast}$ exploits the fundamental symmetry properties of the photon trajectory. The coordinate transformation:
\begin{equation} \label{eq:phi_shift}
    \phi = \frac{\pi}{2} + \psi,
\end{equation}

where $\psi$ denotes the small angular deviation from the straight-line trajectory, enables the identification of the total deflection angle:
\begin{equation} \label{eq:defl_angle}
    \alpha_{\circledast} = 2\psi.
\end{equation}

Through systematic expansion to second order in the gravitational strength parameter $\varepsilon = M u_\bot = M/\gamma$, the analytical expression for the deflection angle becomes:
\begin{equation} \label{islens}
\alpha_{\circledast} \approx \frac{4 M}{\gamma \delta \beta}+\frac{15 M^2 \pi}{4 \gamma^2 \delta^2 \beta^2 }.
\end{equation}

This fundamental result explicitly demonstrates the novel theoretical contribution introduced by the combined LV and CS modifications. The leading-order term exhibits the characteristic $\gamma^{-1}$ scaling modified by the factor $(\delta \beta)^{-1}$, while the second-order correction introduces a $\gamma^{-2}$ contribution absent in standard Schwarzschild lensing \cite{is40}.

The reformulation in terms of the closest approach distance $r_c$ provides enhanced physical insight into the gravitational lensing mechanism. The relationship between the perturbative parameter and closest approach distance is:
\begin{equation} \label{eq:r0}
\frac{1}{r_c} =u_\bot+u_\bot^2 M+\frac{3}{16} u_\bot^3 M^2.
\end{equation}

Substituting this relation yields the deflection angle in the physically meaningful closest approach formulation:
\begin{equation} 
\alpha_{\circledast} \approx \frac{4 M}{\delta \beta r_c}+\frac{M^2}{\delta^2 \beta^2 r_c^2}\left(\frac{15}{4} \pi-4\right). \label{isrmin}
\end{equation}

\begin{figure}[H]
  \centering
      \subfloat[]{\includegraphics[width=0.5\linewidth]{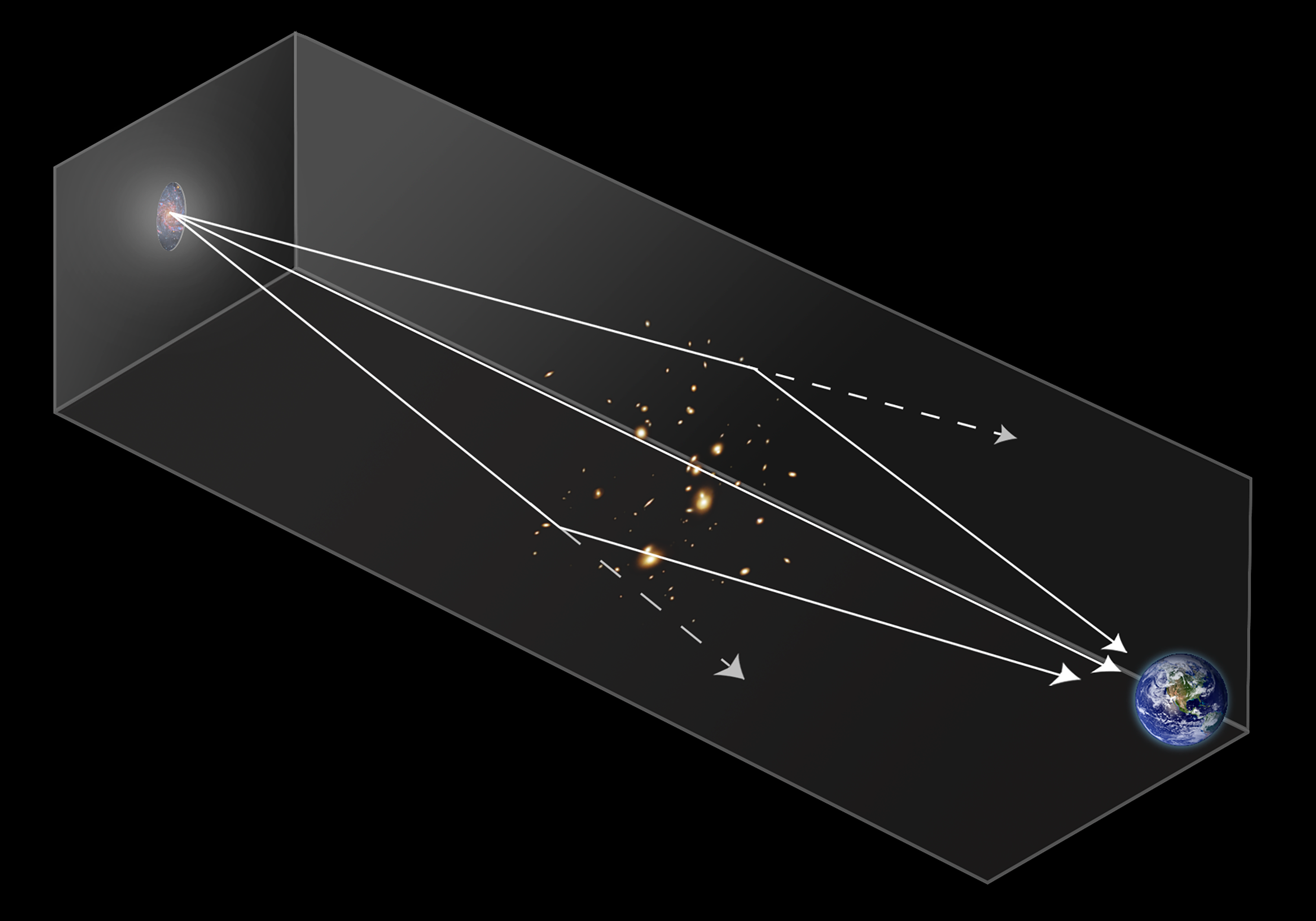}} \\
  \subfloat[]{\includegraphics[width=0.65\linewidth]{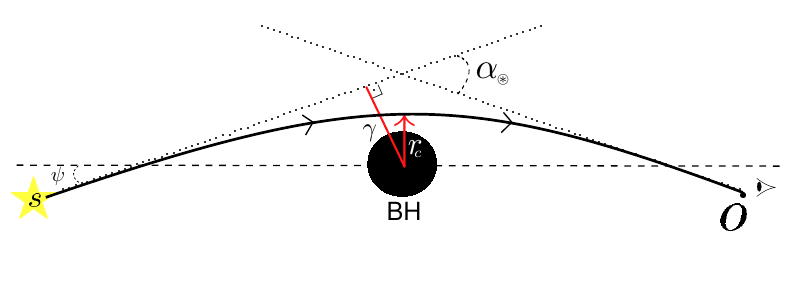}}
  \caption{(a) Gravitational lensing visualization depicting how light from distant galaxies is distorted by an intervening massive object, creating multiple images and extended arcs. This phenomenon emerges directly from the spacetime curvature described by the deflection angle in Eq.~\eqref{isrmin} (Credit: Webbtelescope.org \cite{sec2is16x}). (b) Schematic representation of the SKRCS BH lensing geometry, illustrating the deflection angle $\alpha_{def}$, impact parameter $\gamma$, and the angular separations between source, observer, and resulting images.}
  \label{izfig0}
\end{figure}

This theoretical framework establishes quantitative predictions for gravitational lensing observations in SKRCS spacetimes. The closest approach distance $r_{c}$ represents a directly observable quantity that can be related to experimental measurements, providing enhanced connection between theoretical predictions and observational data compared to the asymptotic impact parameter $\gamma$ \cite{is38}.

The systematic contributions of the LV parameter $\ell$ and CS parameter $\beta$ become particularly pronounced for photon trajectories passing in close proximity to the BH, where higher-order terms in the deflection angle expansion gain substantial importance. Figure~\ref{izfig0} demonstrates the fundamental theoretical connection between SKRCS spacetime modifications and observable gravitational lensing phenomena, where panel (a) illustrates the characteristic light ray distortion patterns that emerge from the modified deflection angle expressions derived in this analysis. The schematic representation in panel (b) systematically depicts the geometric configuration of SKRCS BH lensing, explicitly showing how the theoretical parameters $\ell$, $\beta$, and the resulting deflection angle $\alpha_{\circledast}$ translate into measurable angular separations between lensed images. These visualization frameworks establish the crucial bridge between abstract theoretical formulations and potential observational signatures that could distinguish SKRCS geometries from conventional Schwarzschild spacetimes through precision astrometric measurements. 

\subsection{Deflection Angle of SKRCS BH via Analytical Method} \label{subsec2.2}

The theoretical investigation of exact deflection angles in SKRCS geometries necessitates sophisticated analytical methodologies that transcend perturbative approximations. This comprehensive approach employs elliptic integral formulations to obtain precise analytical expressions for photon bending angles, providing complete mathematical descriptions of gravitational lensing phenomena in modified spacetime geometries \cite{is36,is37}.

The analytical framework establishes solid physical foundations for understanding the intricate interplay between LV mechanisms and CS topological configurations in determining photon trajectory characteristics. Unlike perturbative methodologies that provide approximate solutions, the elliptic integral approach yields exact analytical representations valid across the entire parameter space of SKRCS configurations \cite{is22}.

The exact analytical treatment commences with the fundamental null geodesic equation for equatorial motion $\theta = \pi/2$, as established in the orbit equation \eqref{b10}:
\begin{equation}
\left(\frac{dr}{d\phi}\right)^2 = \beta^2 r^4 \left[ \frac{1}{\gamma^2} - \frac{1}{r^2} \left(\delta - \frac{2M}{\beta r} \right) \right]. \label{iszx1}
\end{equation}

The implementation of the canonical transformation $u = 1/r$ systematically converts this expression into a more mathematically tractable form. This transformation, widely employed in classical gravitational lensing theory \cite{aa2}, yields:
\begin{equation}
\left( \frac{du}{d\phi} \right)^2 = \beta^2 \left( \delta u^2 - \frac{2M}{\beta} u^3 - \frac{1}{\gamma^2} \right).
\end{equation}

Through systematic algebraic manipulation involving multiplication by $\frac{1}{\delta}$ and subsequent regrouping of terms, this equation transforms into the canonical cubic polynomial form:
\begin{equation}
\left( \frac{du}{d\phi} \right)^2 = 2M \beta \left( u - u_1 \right) \left( u_2 - u \right) \left( u_3 - u \right),
\end{equation}

where $u_1 < 0 < u_2 < u_3$ represent the three real roots of the underlying cubic polynomial, with $u_2 = 1/P$ defining the inverse periastron distance parameter.

The analytical framework introduces the periastron parameter $P$ and auxiliary quantity $R$ through the fundamental relations \cite{is26}:
\begin{equation}
u_2 = \frac{1}{P}, \qquad R = P \delta. \label{isn1}
\end{equation}

The systematic matching of coefficients between the factored and expanded cubic polynomial forms establishes explicit expressions for all three roots:
\begin{align} \label{isn2}
u_1 &= \frac{R - 2M - Q}{4MP}, \\
u_2 &= \frac{1}{P}, \\
u_3 &= \frac{R - 2M + Q}{4MP},
\end{align}

where the discriminant quantity $Q$ satisfies:
\begin{equation}
Q^2 = (R - 2M)(R + 6M).
\end{equation}

This cubic root structure systematically incorporates both LV contributions through $\delta = 1/(1 - \ell)$ and CS effects through the parameter $P$, establishing the complete mathematical foundation for exact deflection angle calculations.

With the cubic polynomial completely factorized, the exact deflection angle assumes the integral representation:
\begin{equation}
\hat{\alpha} = 2 \int_0^{u_2} \frac{du}{\sqrt{2M \beta (u - u_1)(u_2 - u)(u_3 - u)}} - \pi.
\end{equation}

The extraction of constant factors yields the standardized form:
\begin{equation}
\hat{\alpha} = \sqrt{\frac{2}{M\beta}} \int_0^{u_2} \frac{du}{\sqrt{(u - u_1)(u_2 - u)(u_3 - u)}} - \pi.
\end{equation}

The systematic decomposition of this integral into constituent elliptic components requires careful treatment of the integration limits and singularity structure. The integral splits into two fundamental elliptic parts:
\begin{equation}
\hat{\alpha} = \sqrt{\frac{2}{M\beta}} \left[ \int_{u_1}^{u_2} \frac{du}{\sqrt{(u - u_1)(u_2 - u)(u_3 - u)}} - \int_{u_1}^{0} \frac{du}{\sqrt{(u - u_1)(u_2 - u)(u_3 - u)}} \right] - \pi.
\end{equation}

The analytical evaluation of these integrals employs the theory of incomplete elliptic integrals of the first kind $F(\Psi, k)$, providing exact closed-form expressions \cite{sec2is23}. The complete analytical result takes the form:
\begin{equation}
\hat{\alpha} = \sqrt{\frac{2}{M\beta}} \left[ \frac{2 F(\Psi_1, k)}{\sqrt{u_3 - u_1}} - \frac{2 F(\Psi_2, k)}{\sqrt{u_3 - u_1}} \right] - \pi,
\end{equation}

where the elliptic integral parameters are systematically determined by the cubic root structure. The complete elliptic integral $K(k) = F(\pi/2, k)$ and incomplete elliptic integral $F(\Psi_2, k)$ are characterized by the amplitude and modulus parameters \cite{sec2is24}:
\begin{align}
\Psi_1 &= \frac{\pi}{2}, \\
\Psi_2 &= \sin^{-1} \sqrt{\frac{-u_1}{u_2 - u_1}}, \\
k^2 &= \frac{u_2 - u_1}{u_3 - u_1}.
\end{align}

The elliptic integral formulation enables alternative parametric representations that enhance computational efficiency and theoretical insight. The amplitude parameter $\Psi_2$ admits the equivalent expression:
\begin{equation}
\Psi_2 = \sin^{-1} \sqrt{\frac{Q + 2M - R}{Q + 6M - R}}.
\end{equation}

The theoretical framework exhibits elegant limiting behavior that recovers established results in appropriate parameter regimes. The simultaneous limits $\delta \to 1$ and $\beta \to 1$ systematically reproduce the classical Schwarzschild deflection angle \cite{is41}, confirming the mathematical consistency of the analytical formulation with established gravitational lensing theory. The elliptic integral representation provides computationally efficient pathways for numerical evaluation across the complete SKRCS parameter space. This exact analytical framework establishes theoretical foundations for investigating observable signatures of LV and CS modifications in precision gravitational lensing measurements. Namely, they could offer observational pathways for constraining exotic physics beyond conventional GR through astronomical observations of deflection angle variations in different astrophysical environments.

\section{Magnification of SKRCS BHs} \label{isec3}

The gravitational lensing phenomenon represents one of the most profound manifestations of spacetime curvature, wherein the magnification effects induced by SKRCS BHs provide unprecedented insights into the fundamental nature of modified gravitational theories. This comprehensive investigation systematically analyzes how magnification properties undergo systematic modifications due to combined LV and CS deformation parameters, substantially extending beyond conventional Schwarzschild lensing frameworks through the incorporation of exotic gravitational physics \cite{sec2is27,sec2is28}.

The theoretical exploration of SKRCS magnification phenomena necessitates sophisticated analytical methodologies that systematically account for the intricate interplay between topological defects and LV mechanisms in determining observable lensing signatures. These modified spacetime geometries introduce novel parameter dependencies that fundamentally alter the magnification characteristics, potentially offering observational pathways for constraining theoretical deviations from standard GR predictions \cite{sec2is29,sec2is29x}.

The geometric foundation of SKRCS lensing establishes the relationship between source and image angular positions through the deflection angle formalism developed in previous sections. The spherical symmetry inherent in SKRCS geometries ensures that gravitational lensing affects exclusively radial distance measurements while preserving azimuthal angle invariance, leading to the fundamental lens equation \cite{sec2is30}:

\begin{equation}
({\Gamma} - {\theta}) D_{\rm s} =\alpha_{\circledast} D_{\rm ls}
\end{equation}

This relationship can be systematically reformulated to yield the explicit dependence of image position $\Gamma$ on source position $\theta$:

\begin{equation}
{\Gamma} = {\theta}+\frac{D_{\rm {ls}}}{D_{\rm s}}
\alpha_{\circledast}\,\,.
\end{equation}

\begin{figure}[ht!]
   \centering
       \subfloat[]{\includegraphics[width=0.5\linewidth]{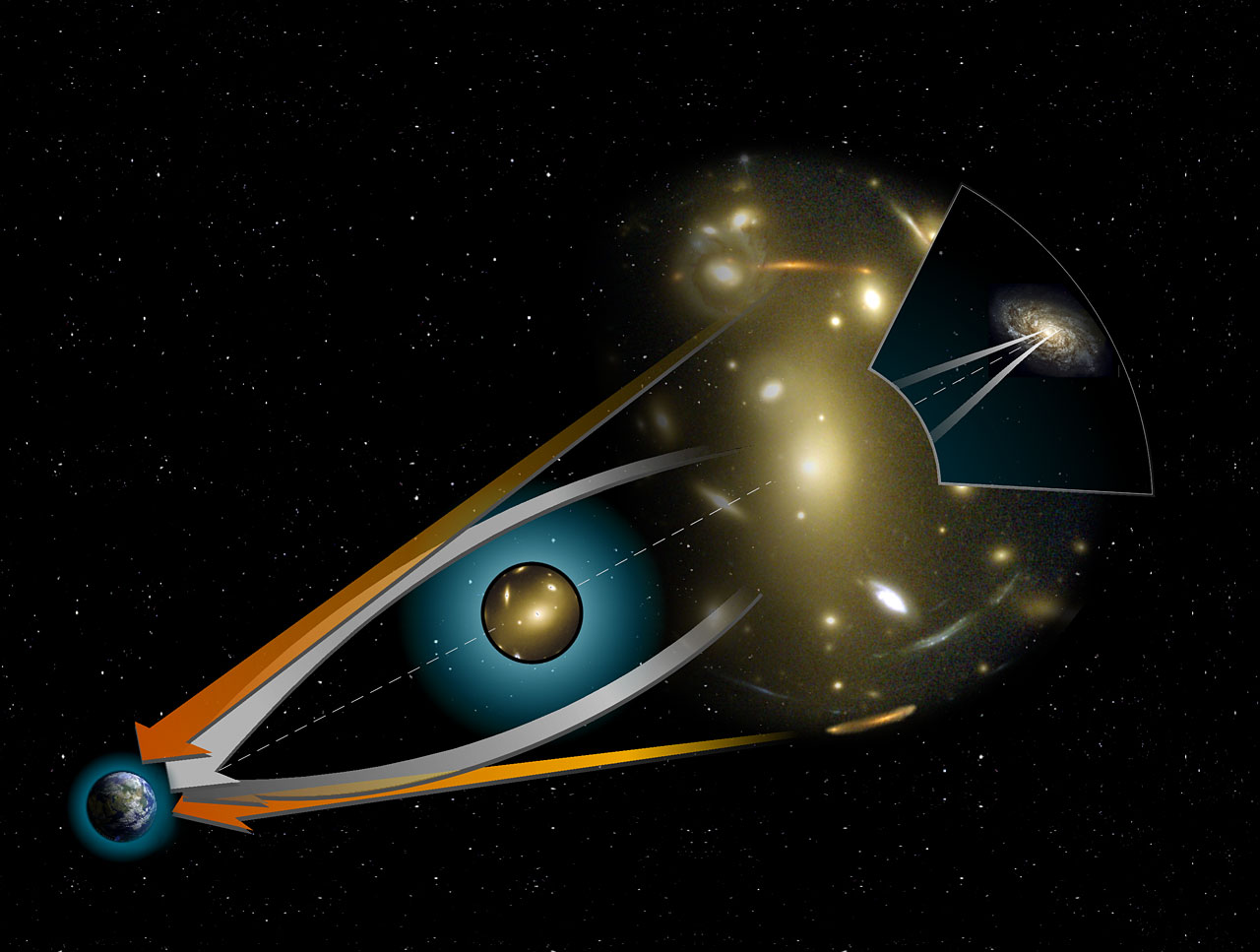}} \\
   \subfloat[]{\includegraphics[width=0.3\linewidth]{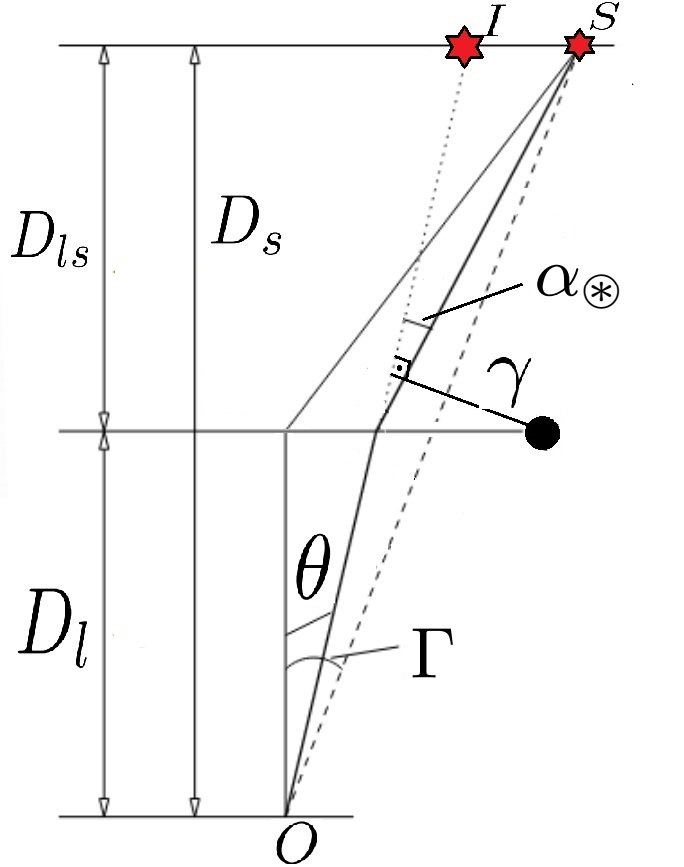}}
   \caption{(a) Hubble's observations illustrate how the gravity of a foreground BH bends and deflects the light from a distant star (Credit: Science.nasa.gov \cite{sec2is16}). (b) Lensing illustration of a BH: $O$ is the observer, $S$ is the source, BH is the BH acting as a gravitational lens, $I$ is the position of the image. $\beta$ is the angle between the source and the optical axis, $\theta$ is the angle between the image and the optical axis, $\alpha_{\text{def}}$ is the deflection angle. $b$ is the impact parameter and $D_l$, $D_s$ and $D_{ls}$ are angular diameter distances (from observer to lens, from observer to source, and from lens to source, respectively). The dashed line represents the path light would take in the absence of the lens, while the solid line shows the actual path of light due to gravitational lensing.}
   \label{izo1}
\end{figure}

The transition from theoretical natural units to observationally relevant SI units requires systematic incorporation of fundamental physical constants, specifically the gravitational constant $G$ and the speed of light $c$. This transformation, implemented through $M \rightarrow M \cdot G/c^2$, enables direct quantitative comparison with astronomical observations and facilitates integration with existing observational databases from precision astrometric surveys \cite{sec2is32}.

Incorporating the previously derived deflection angle from Eq.~\eqref{islens}, the comprehensive lens equation for SKRCS BH configurations becomes:

\begin{equation}
\Gamma = \theta +{\bar{\theta}_{\xi}^2}{\theta}^{-1},
\end{equation}

where the modified Einstein angle incorporates both LV and CS contributions:

\begin{equation}
\bar{\theta}_{\xi}^2 = {\theta}_{\xi}^2\left(\frac{1}{\delta\beta}+\frac{15 \pi G M}{16 \theta c^2 \delta^2 \beta^2 D_l}\right),
\end{equation}

with the standard Einstein angle defined as:
\begin{equation}
\theta_{\xi}^2=4MG/c^2 \times \frac{D_{\mathrm{ls}}}{D_{\mathrm{s}} D_l}.
\end{equation}

Contemporary theoretical developments have established that gravitational lensing phenomena extend beyond mere directional modifications to encompass systematic alterations in the cross-sectional area of photon ray bundles. These cross-sectional modifications directly influence the observed brightness characteristics of lensed astronomical objects, quantified through sophisticated magnification calculation methodologies \cite{sec2is33}. 

For infinitesimally small source configurations, the total magnification factor $\mu_{tot}$ is determined by:

\begin{equation}
\mu^{-1} = \left|\frac{\Gamma}{\theta}\frac{d\Gamma}{d\theta}\right|\,\,.
\end{equation}

This formulation demonstrates that lensed images undergo magnification or demagnification by the factor $|\mu|$, with multiple image formation scenarios requiring summation of individual image magnification contributions to determine total brightness amplification effects.

The systematic decomposition of magnification effects into tangential and radial components provides fundamental insight into the geometric structure of gravitational lensing phenomena. These components represent mathematically distinct contributions to the overall magnification, characterized by divergent behavior at critical curves where magnification theoretically approaches infinity.

The tangential magnification component is expressed as:

\begin{equation}
\mu_{\rm tan}=\left|\frac{\Gamma}{\theta}\right|^{-1}
=\theta^2\left(\theta^2+\tilde{\theta}_{\xi}^2\right)^{-1},
\end{equation}

while the radial magnification exhibits:
\begin{equation}
\mu_{\rm rad} = \left|\frac{d\Gamma}{d\theta}\right|^{-1} =\theta^2\left(\theta^2-\tilde{\theta}_{\xi}^2\right)^{-1}.
\end{equation}
A fundamental characteristic emerges through analysis of these magnification components: $\mu_{\rm rad}$ exhibits mathematical singularities at $\theta=\theta_{\rm E}$, corresponding to the angular radius of radial critical curves, while $\mu_{\rm tan}$ maintains finite values throughout the accessible parameter space. This distinctive behavior suggests that SKRCS BH configurations generate unique observational signatures that potentially enable differentiation from alternative compact object models through precision astronomical observations.
\vspace{-4mm}
\noindent\begin{figure}[H]
\centering
\begin{tikzpicture}[inner sep=0pt, outer sep=0pt]

\draw[thick] (0.5\linewidth,0) -- (0.5\linewidth,-0.9\linewidth);
\draw[thick] (0,-0.45\linewidth) -- (\linewidth,-0.45\linewidth);

\node[anchor=north west] at (0,0){
  \begin{minipage}[t][0.45\linewidth][c]{0.45\linewidth}
    \centering
    \includegraphics[width=0.8\linewidth]{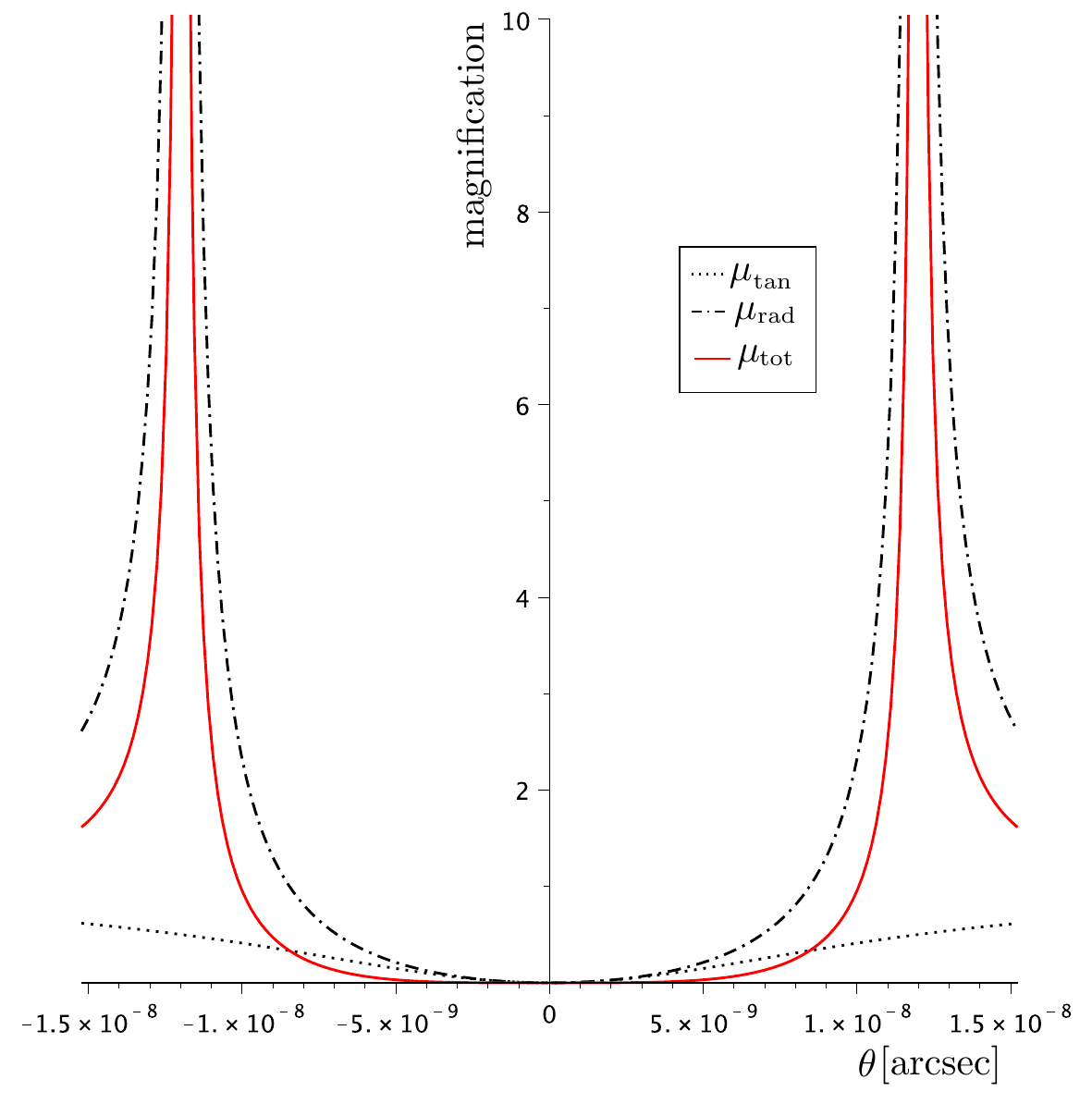}
    
    (a) $\ell = 0.06092$ \\
    \quad {\small\textit{(Within Milky Way Galaxy Bound: $-0.18502<\ell<0.06093$)}}
  \end{minipage}
};

\node[anchor=north west] at (0.5\linewidth,0){
  \begin{minipage}[t][0.45\linewidth][c]{0.45\linewidth}
    \centering
    \includegraphics[width=0.8\linewidth]{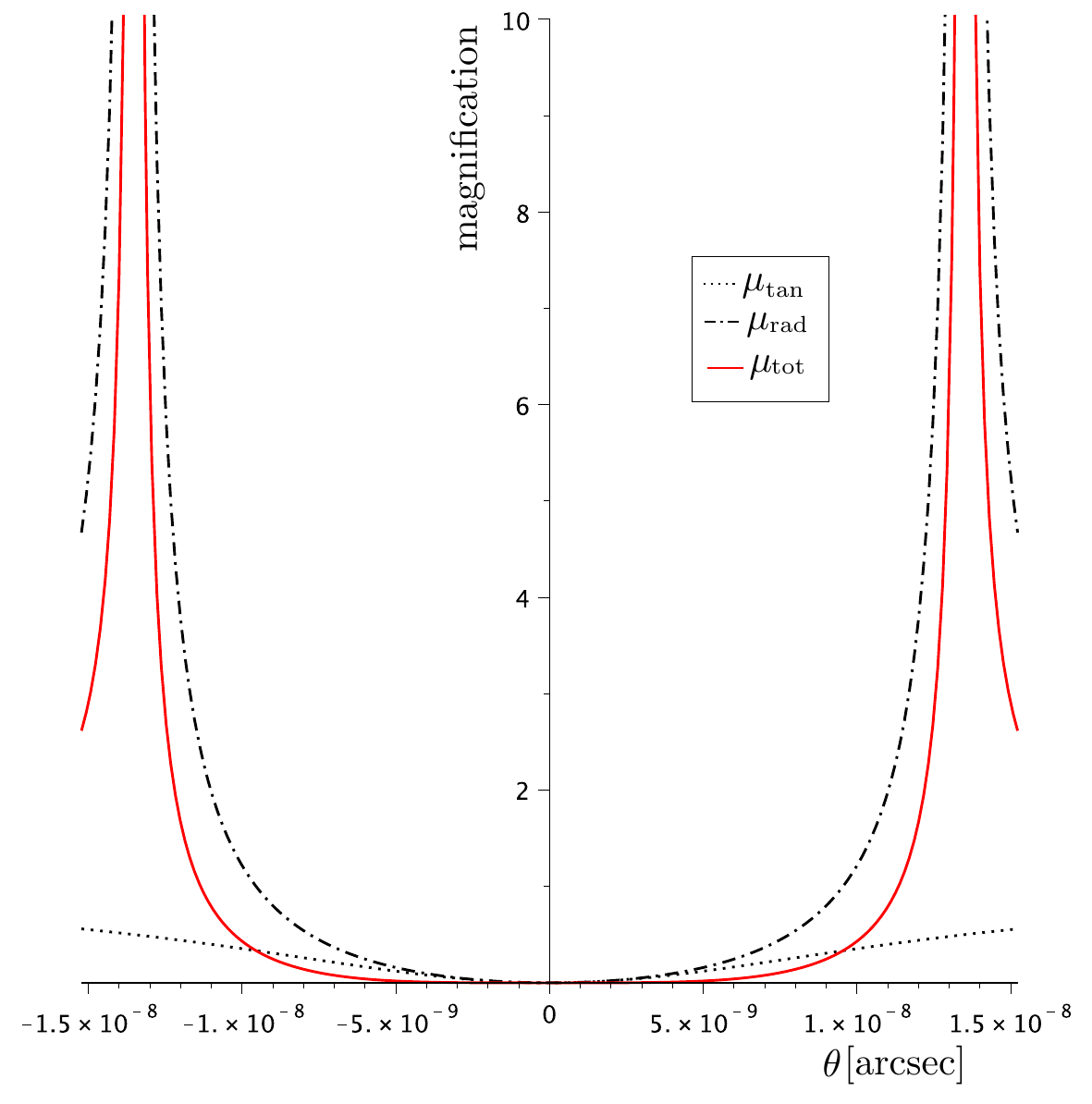}
    
    (b) $\ell = -0.18501$ \\
    \quad {\small\textit{(Within Milky Way Galaxy Bound: $-0.18502<\ell<0.06093$)}}
  \end{minipage}
};

\node[anchor=north west] at (0,-0.45\linewidth){
  \begin{minipage}[t][0.45\linewidth][c]{0.45\linewidth}
    \centering
    \includegraphics[width=0.8\linewidth]{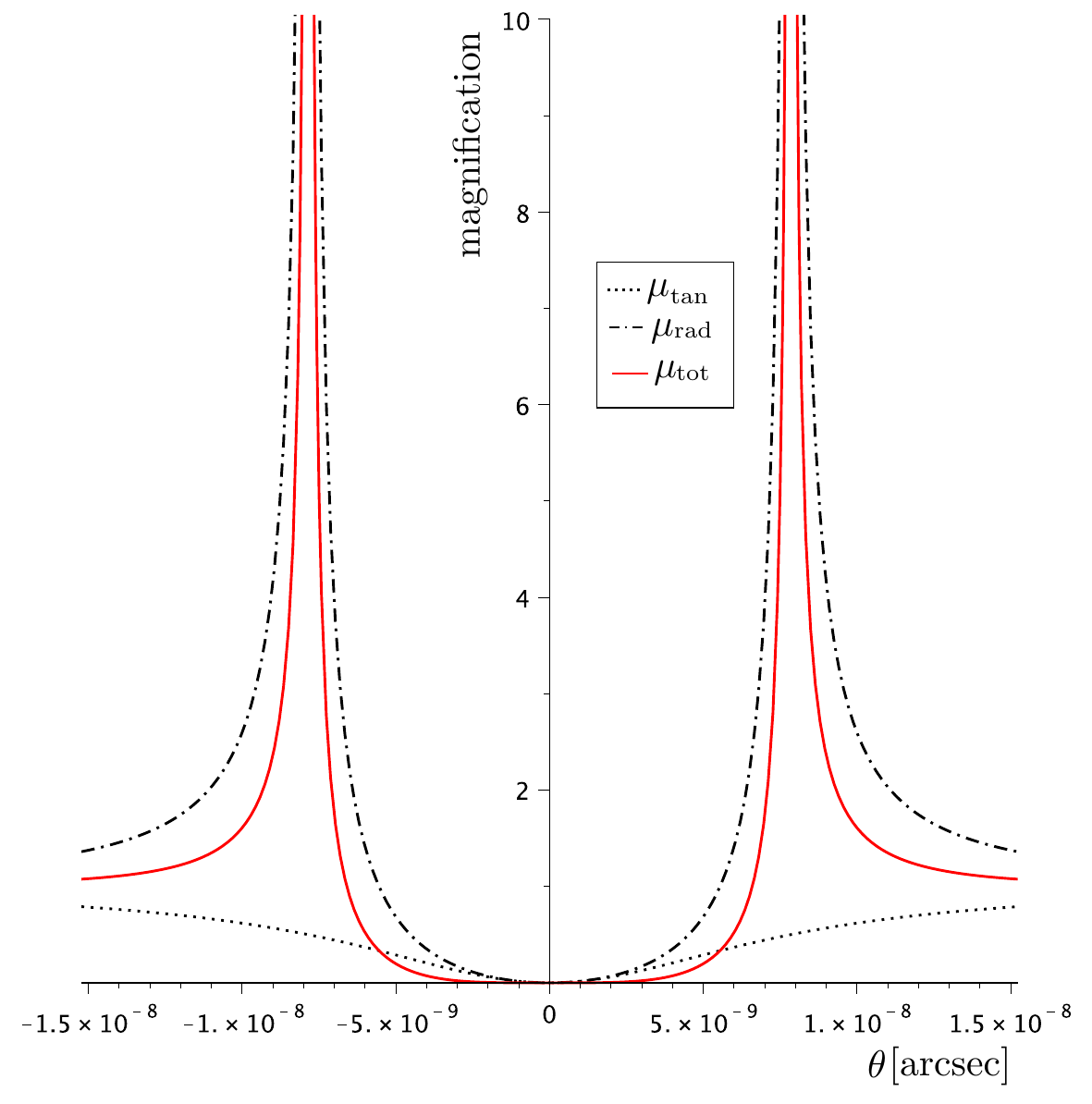}
    
    (c) $\ell = 0.6$\\
    \quad {\small\textit{(Theoretically Acceptable Value)}}
  \end{minipage}
};

\node[anchor=north west] at (0.5\linewidth,-0.45\linewidth){
  \begin{minipage}[t][0.45\linewidth][c]{0.5\linewidth}
    \centering
    \includegraphics[width=0.8\linewidth]{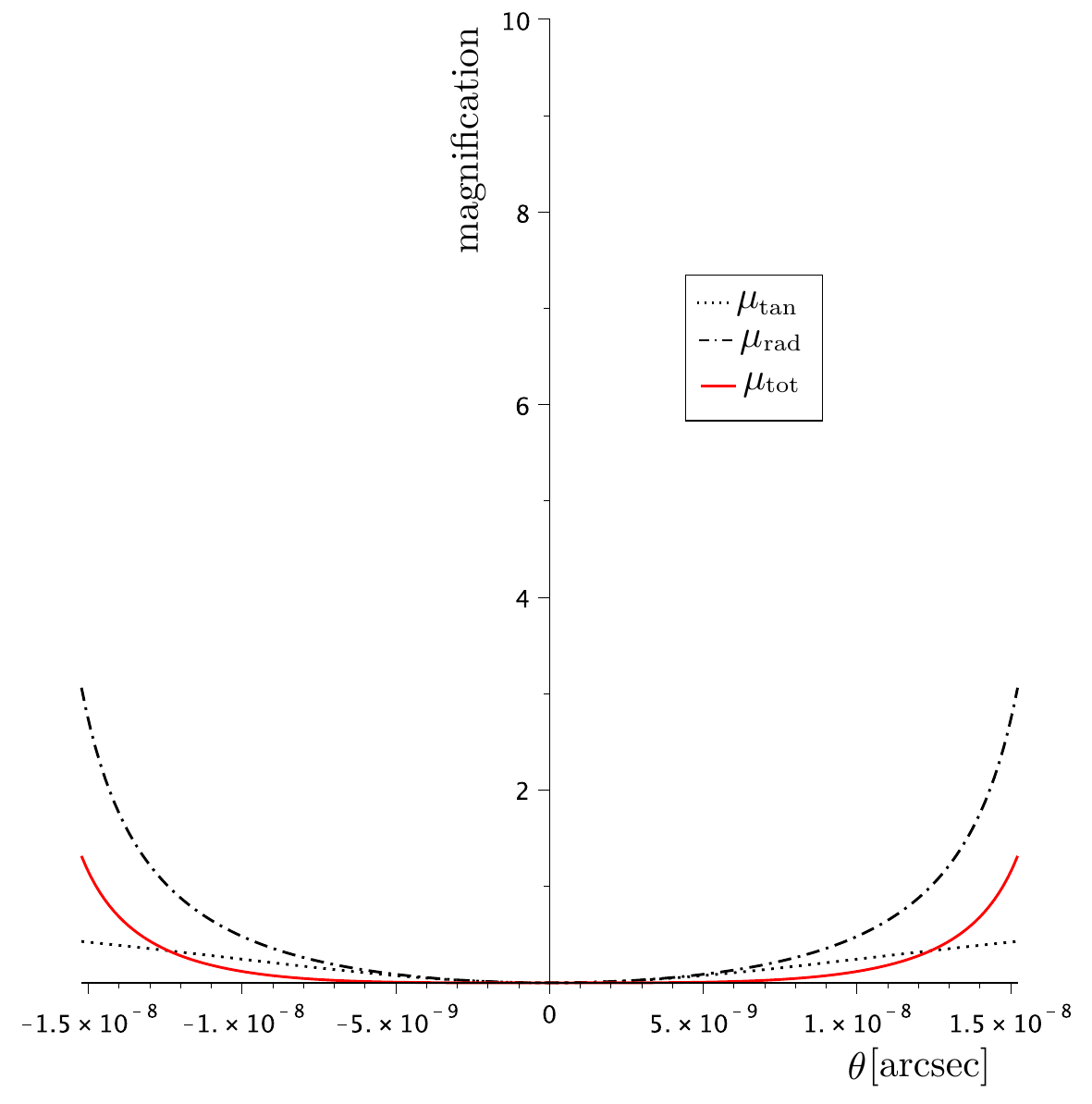}
    
    (d) $\ell = -1$\\
    \quad {\small\textit{(Theoretically Acceptable Value)}}
  \end{minipage}
};

\end{tikzpicture}

\vspace{0.2cm}
\caption{The magnifications—tangential $\mu_{\rm tan}$ (dotted lines), radial $\mu_{\rm rad}$ (dash-dotted lines), and total $\mu$ (continuous curves)—are plotted as functions of the image position $\theta$ for four SKRCS BH cases with $\beta=0.1$. Panels (a)–(d) correspond to different values of $\ell$ as labeled. The singularities of $\mu_{\rm tan}$ and $\mu_{\rm rad}$ give the positions of the tangential and radial critical curves, respectively. Here $|M|=1\, M_{\odot}$, $D_{\rm s}=0.05\text{ Mpc}$ and $D_{\rm l}=0.01\text{ Mpc}$. Angles are in arcseconds: $1\text{ arcsec}=4.848\times10^{-6}\text{ rad}$.}
\label{isfigLENSBETA01}
\end{figure}

Figure~\ref{isfigLENSBETA01} demonstrates the systematic influence of LV parameter variations on magnification characteristics within the strong CS regime $\beta=0.1$, revealing profound modifications to conventional gravitational lensing behavior. The observationally constrained scenarios in panels (a) and (b) exhibit substantially different critical curve positioning and magnification amplitude scaling compared to theoretical exploration regimes in panels (c) and (d). These variations manifest through systematic shifts in both tangential and radial magnification singularity locations, indicating that CS-dominated configurations amplify LV effects on observable lensing signatures. The theoretical parameter space exploration in panels (c) and (d) reveals extreme magnification enhancement scenarios that, while potentially unphysical under current observational constraints, provide crucial theoretical benchmarks for understanding the fundamental scaling relationships between LV and CS contributions to gravitational lensing phenomena. The dramatic differences between positive and negative LV parameter regimes suggest asymmetric gravitational lensing responses that could potentially distinguish between different theoretical implementations of LV mechanisms through precision astronomical observations.

\begin{figure}[H]
\centering
\begin{tikzpicture}[inner sep=0pt, outer sep=0pt]

\draw[thick] (0.5\linewidth,0) -- (0.5\linewidth,-0.9\linewidth);
\draw[thick] (0,-0.45\linewidth) -- (\linewidth,-0.45\linewidth);

\node[anchor=north west] at (0,0){
  \begin{minipage}[t][0.45\linewidth][c]{0.45\linewidth}
    \centering
    \includegraphics[width=0.8\linewidth]{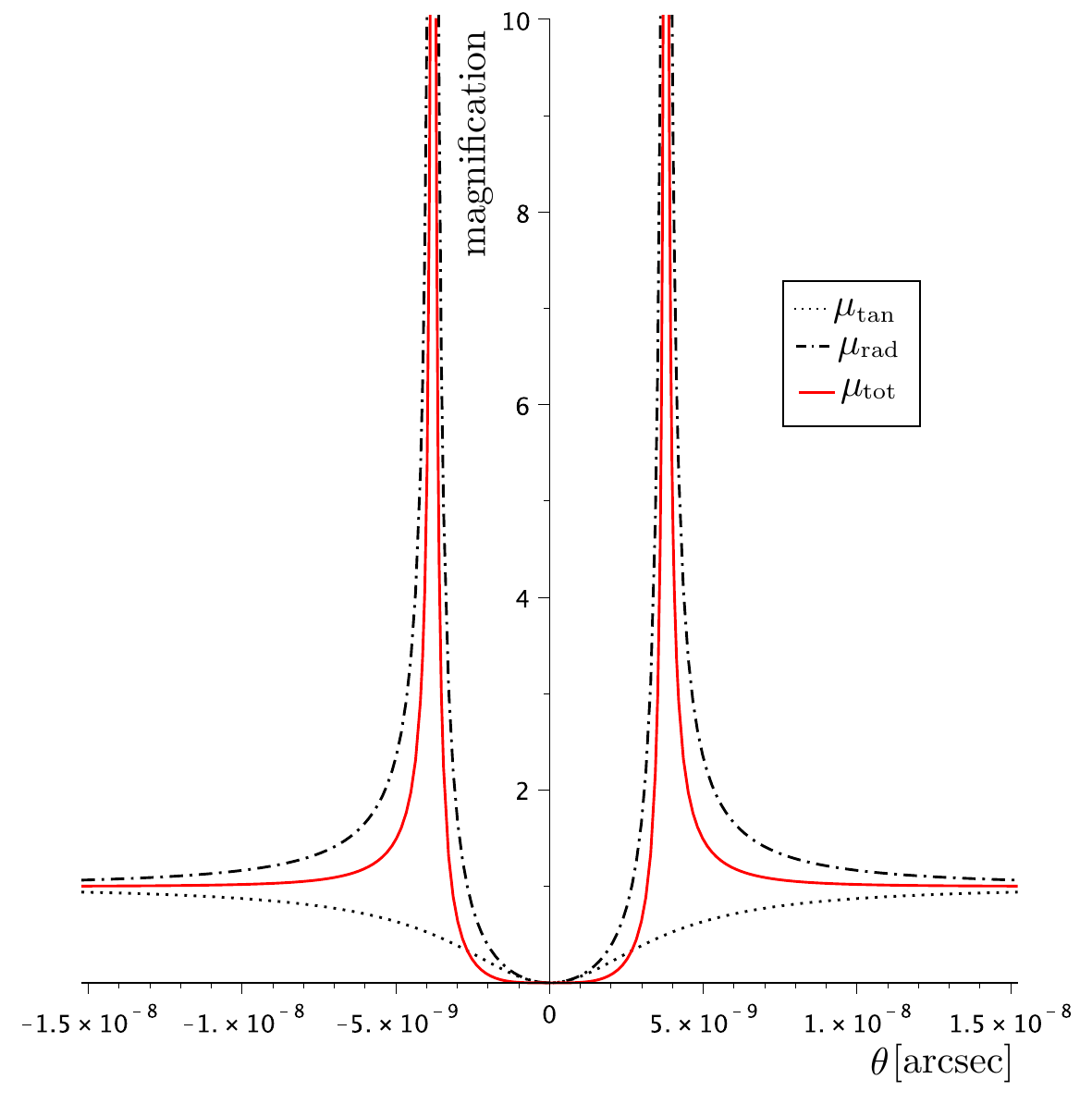}
    
    (a) $\ell = 0.06092$ \\
    \quad {\small\textit{(Within Milky Way Galaxy Bound: $-0.18502<\ell<0.06093$)}}
  \end{minipage}
};

\node[anchor=north west] at (0.5\linewidth,0){
  \begin{minipage}[t][0.45\linewidth][c]{0.45\linewidth}
    \centering
    \includegraphics[width=0.8\linewidth]{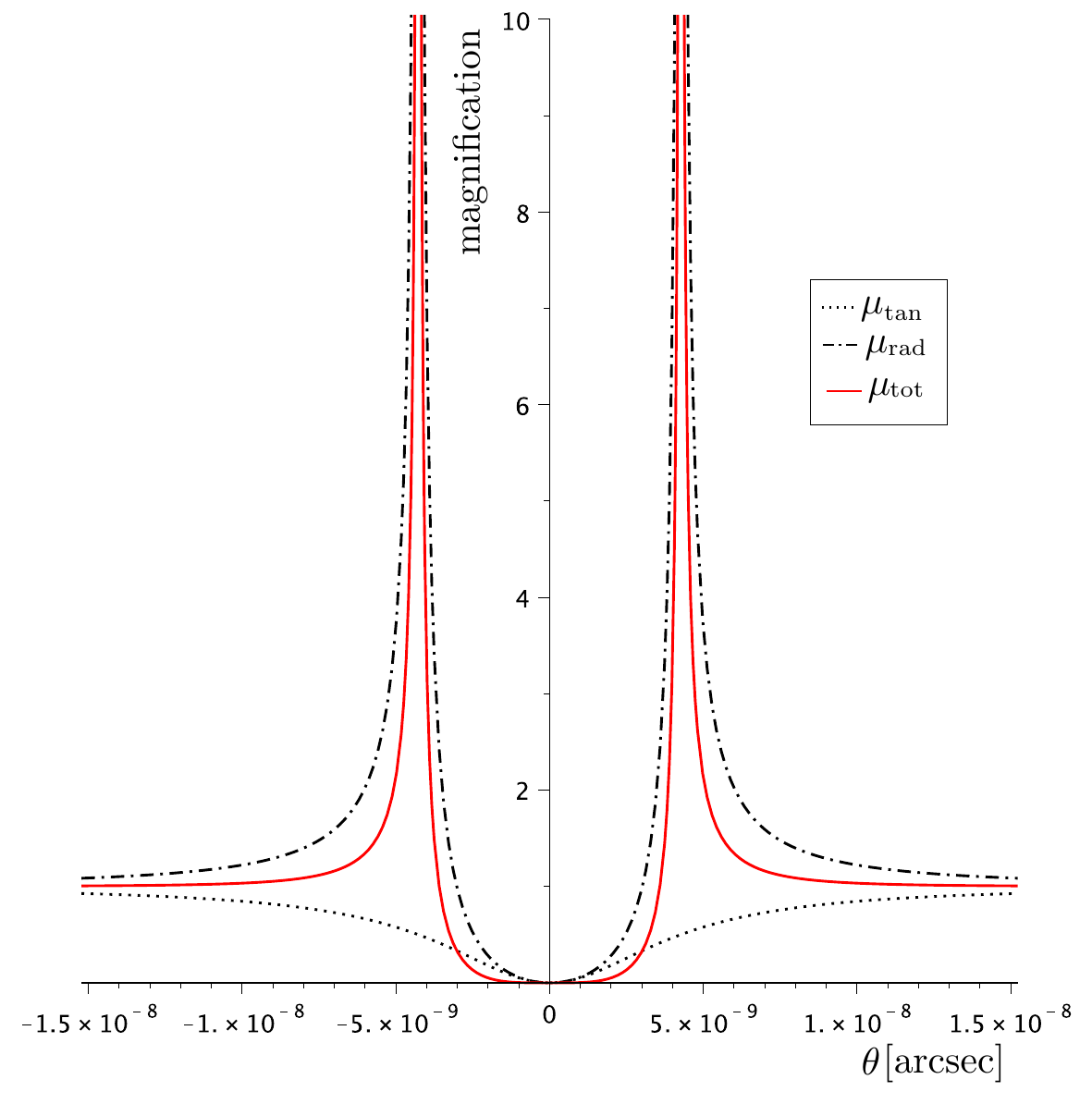}
    
    (b) $\ell = -0.18501$ \\
    \quad {\small\textit{(Within Milky Way Galaxy Bound: $-0.18502<\ell<0.06093$)}}
  \end{minipage}
};

\node[anchor=north west] at (0,-0.45\linewidth){
  \begin{minipage}[t][0.45\linewidth][c]{0.45\linewidth}
    \centering
    \includegraphics[width=0.8\linewidth]{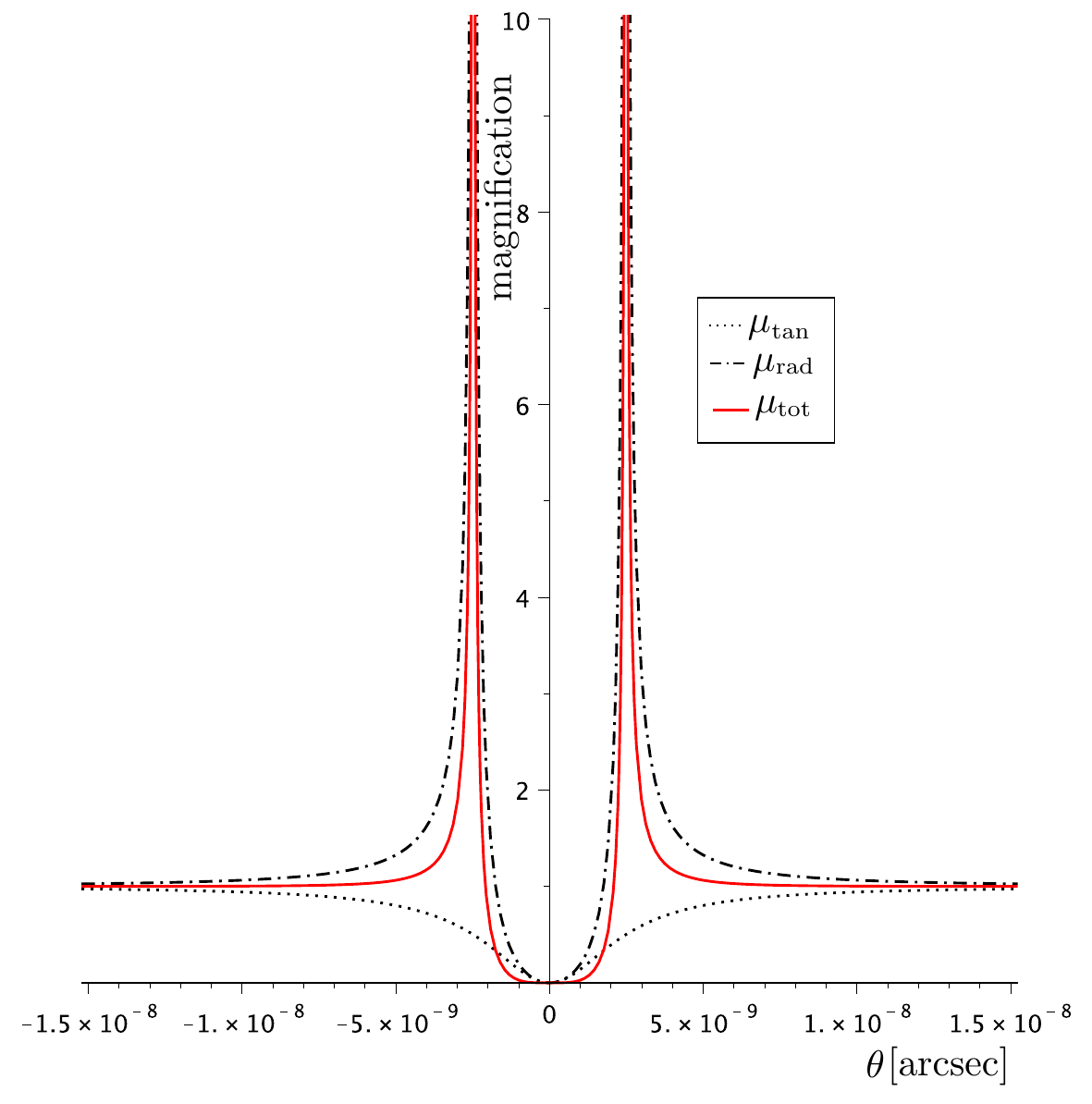}
    
    (c) $\ell = 0.6$\\
    \quad {\small\textit{(Theoretically Acceptable Value)}}
  \end{minipage}
};

\node[anchor=north west] at (0.5\linewidth,-0.45\linewidth){
  \begin{minipage}[t][0.45\linewidth][c]{0.5\linewidth}
    \centering
    \includegraphics[width=0.8\linewidth]{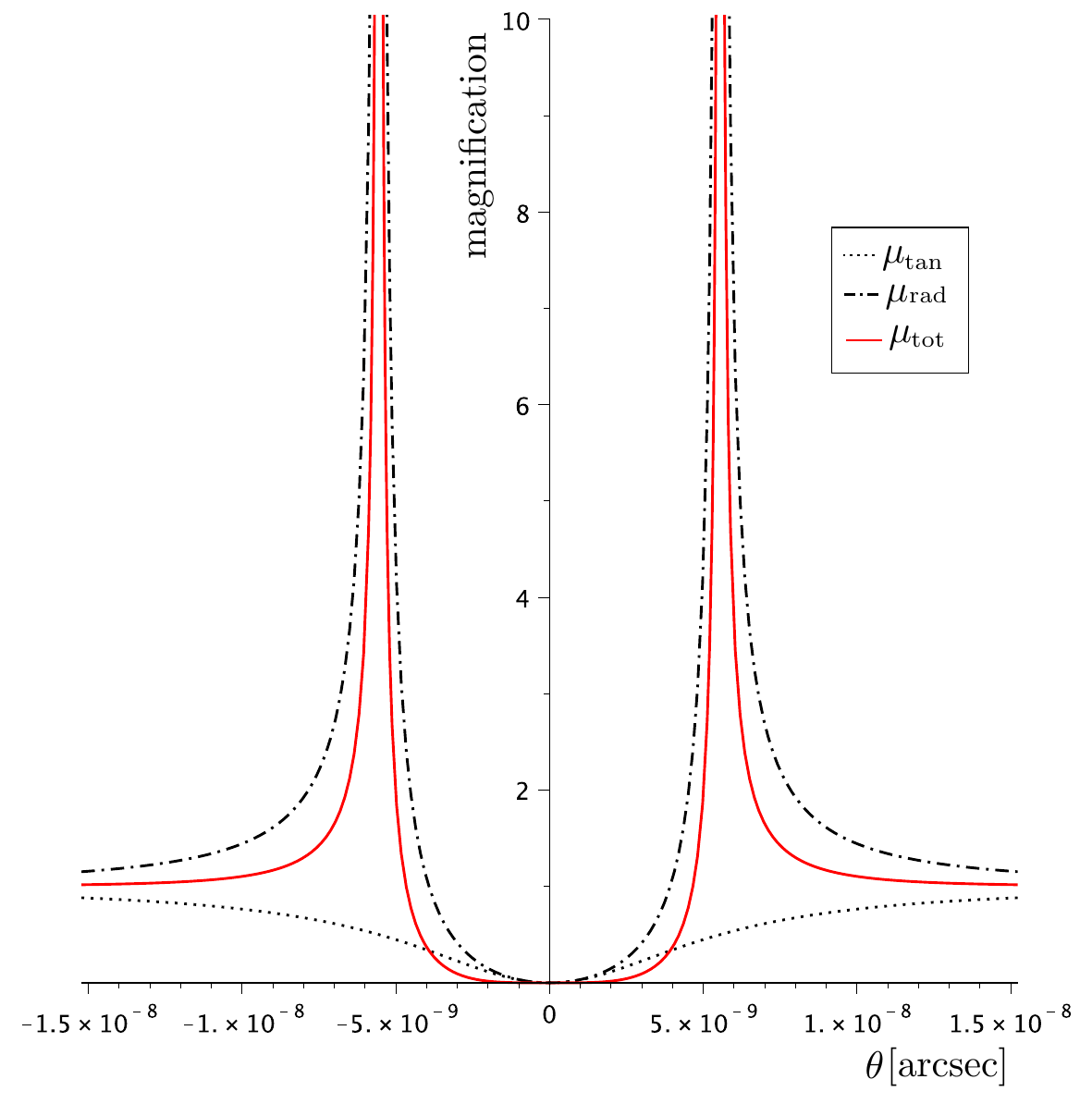}
    
    (d) $\ell = -1$\\
    \quad {\small\textit{(Theoretically Acceptable Value)}}
  \end{minipage}
};

\end{tikzpicture}

\vspace{0.2cm}
\caption{The magnifications—tangential $\mu_{\rm tan}$ (dotted lines), radial $\mu_{\rm rad}$ (dash-dotted lines), and total $\mu$ (continuous curves)—are plotted as functions of the image position $\theta$ for four SKRCS BH cases with $\beta=1$. Panels (a)–(d) correspond to different values of $\ell$ as labeled. The singularities of $\mu_{\rm tan}$ and $\mu_{\rm rad}$ give the positions of the tangential and radial critical curves, respectively. Here $|M|=1\, M_{\odot}$, $D_{\rm s}=0.05\text{ Mpc}$ and $D_{\rm l}=0.01\text{ Mpc}$. Angles are in arcseconds: $1\text{ arcsec}=4.848\times10^{-6}\text{ rad}$.}
\label{isfigLENSBETA1}
\end{figure}

Figure~\ref{isfigLENSBETA1} illustrates the fundamental transition regime where CS effects become negligible ($\beta=1$), effectively reducing the SKRCS configuration to pure KR gravity modifications, thereby providing essential baseline comparisons for understanding CS contributions to gravitational lensing phenomena. The observationally constrained LV parameters in panels (a) and (b) demonstrate relatively modest deviations from standard Schwarzschild lensing behavior, with critical curve positioning shifts that remain within potentially detectable ranges using contemporary precision astrometric techniques. The theoretical exploration regimes in panels (c) and (d) reveal more dramatic magnification modifications that, while exceeding current observational bounds, establish crucial theoretical benchmarks for understanding the fundamental scaling relationships inherent in pure LV gravitational lensing effects. The systematic comparison between positive and negative LV parameter configurations demonstrates pronounced asymmetries in magnification response characteristics, suggesting that gravitational lensing observations could potentially discriminate between different theoretical implementations of LV mechanisms. The elimination of CS contributions in this configuration enables isolation of pure LV effects, providing theoretical foundation for disentangling the relative contributions of topological defects versus fundamental spacetime modifications in future observational programs.

\begin{figure}[H]
\centering
\begin{tikzpicture}[inner sep=0pt, outer sep=0pt]

\draw[thick] (0.5\linewidth,0) -- (0.5\linewidth,-0.9\linewidth);
\draw[thick] (0,-0.45\linewidth) -- (\linewidth,-0.45\linewidth);

\node[anchor=north west] at (0,0){
  \begin{minipage}[t][0.45\linewidth][c]{0.45\linewidth}
    \centering
    \includegraphics[width=0.8\linewidth]{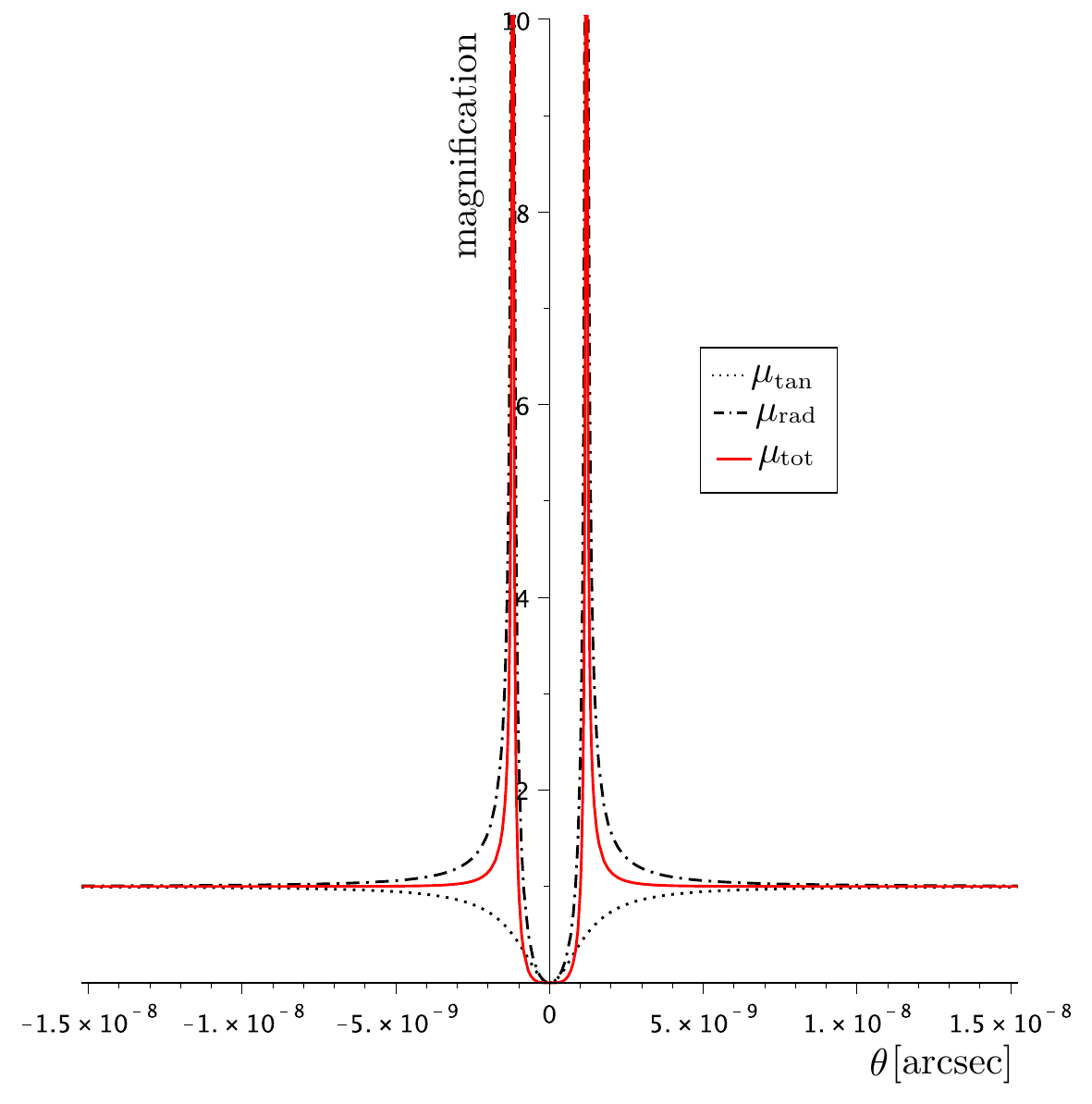}
    
    (a) $\ell = 0.06092$ \\
    \quad {\small\textit{(Within Milky Way Galaxy Bound: $-0.18502<\ell<0.06093$)}}
  \end{minipage}
};

\node[anchor=north west] at (0.5\linewidth,0){
  \begin{minipage}[t][0.45\linewidth][c]{0.45\linewidth}
    \centering
    \includegraphics[width=0.8\linewidth]{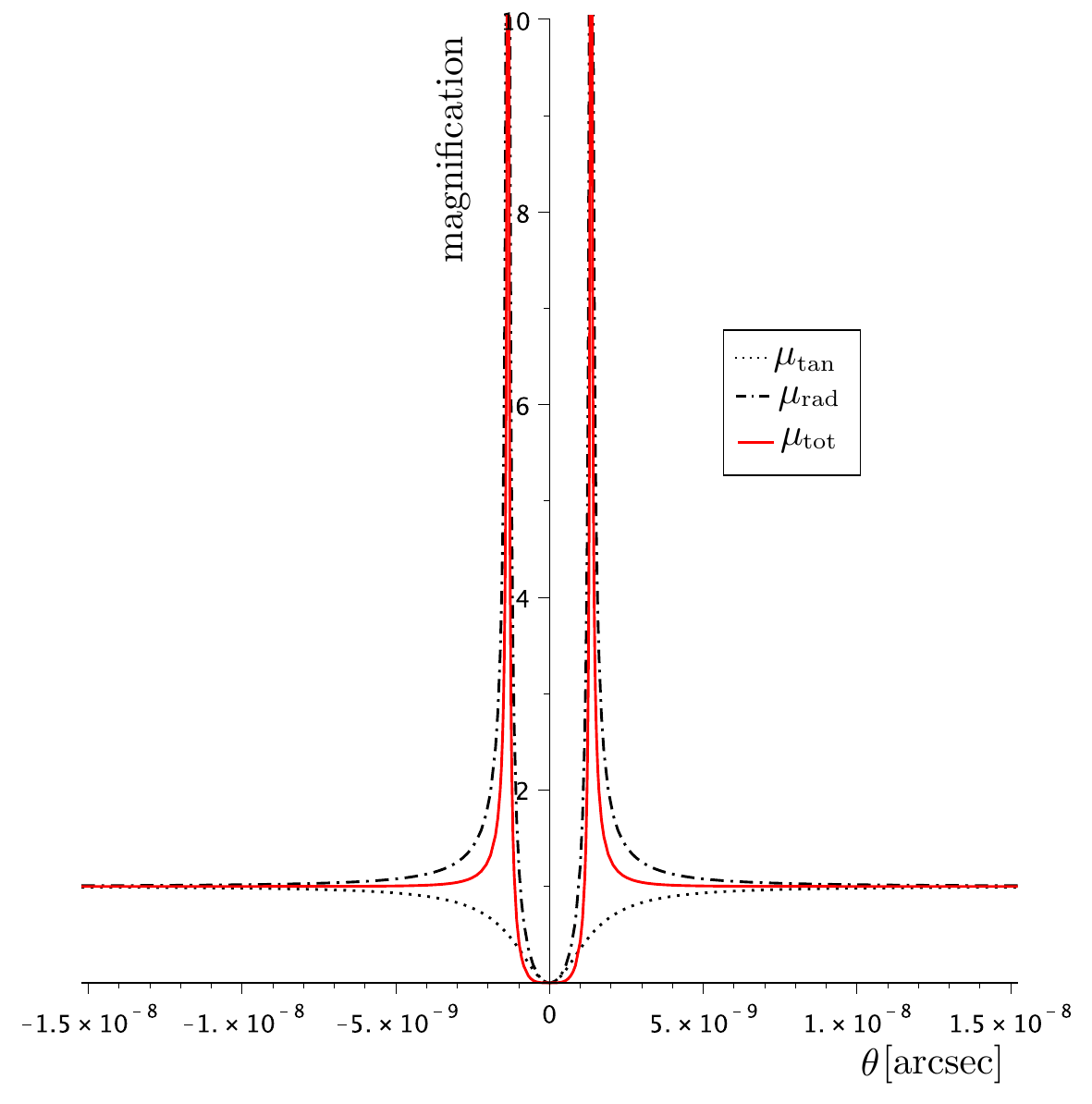}
    
    (b) $\ell = -0.18501$ \\
    \quad {\small\textit{(Within Milky Way Galaxy Bound: $-0.18502<\ell<0.06093$)}}
  \end{minipage}
};

\node[anchor=north west] at (0,-0.45\linewidth){
  \begin{minipage}[t][0.45\linewidth][c]{0.45\linewidth}
    \centering
    \includegraphics[width=0.8\linewidth]{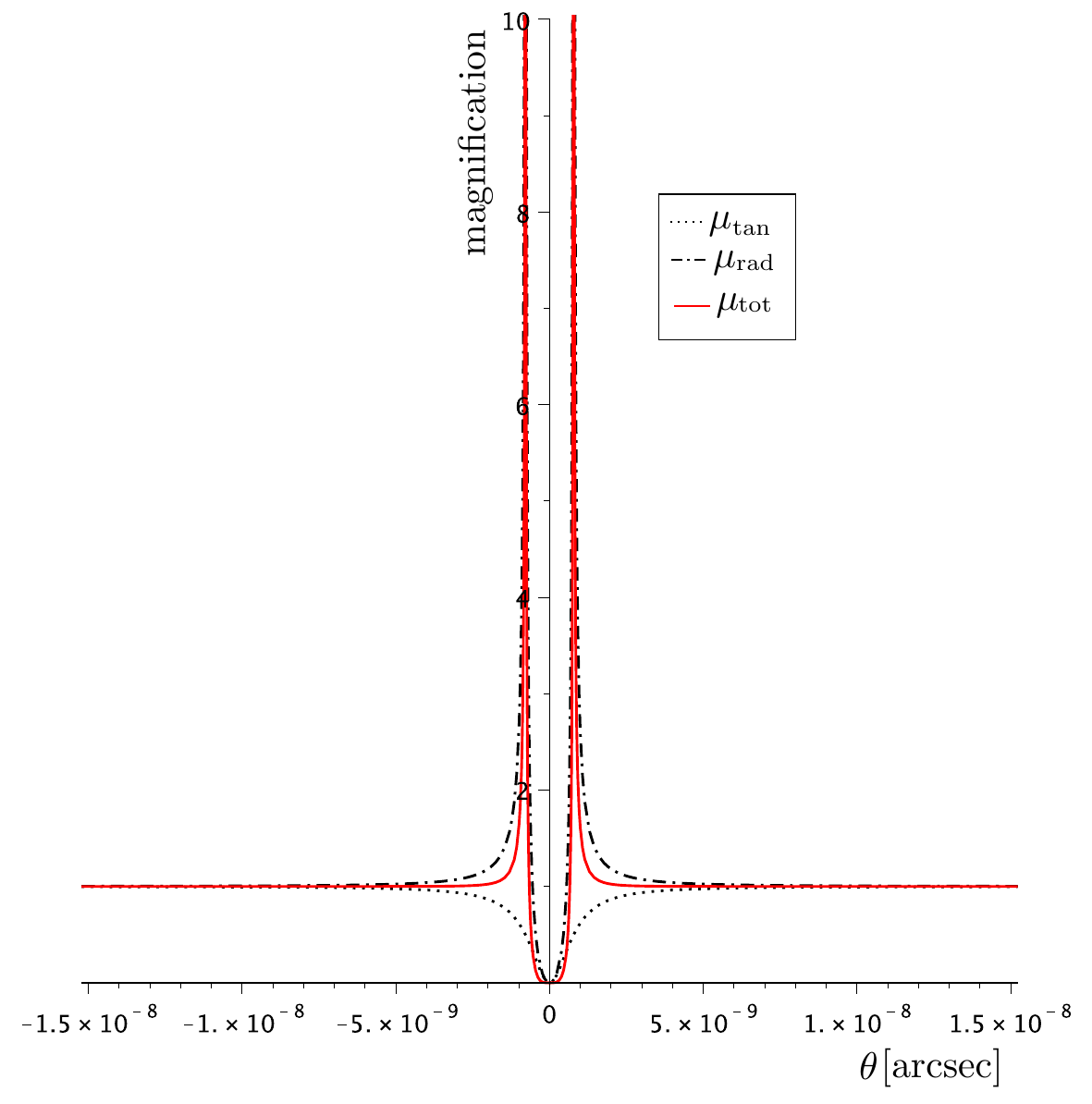}
    
    (c) $\ell = 0.6$\\
    \quad {\small\textit{(Theoretically Acceptable Value)}}
  \end{minipage}
};

\node[anchor=north west] at (0.5\linewidth,-0.45\linewidth){
  \begin{minipage}[t][0.45\linewidth][c]{0.5\linewidth}
    \centering
    \includegraphics[width=0.8\linewidth]{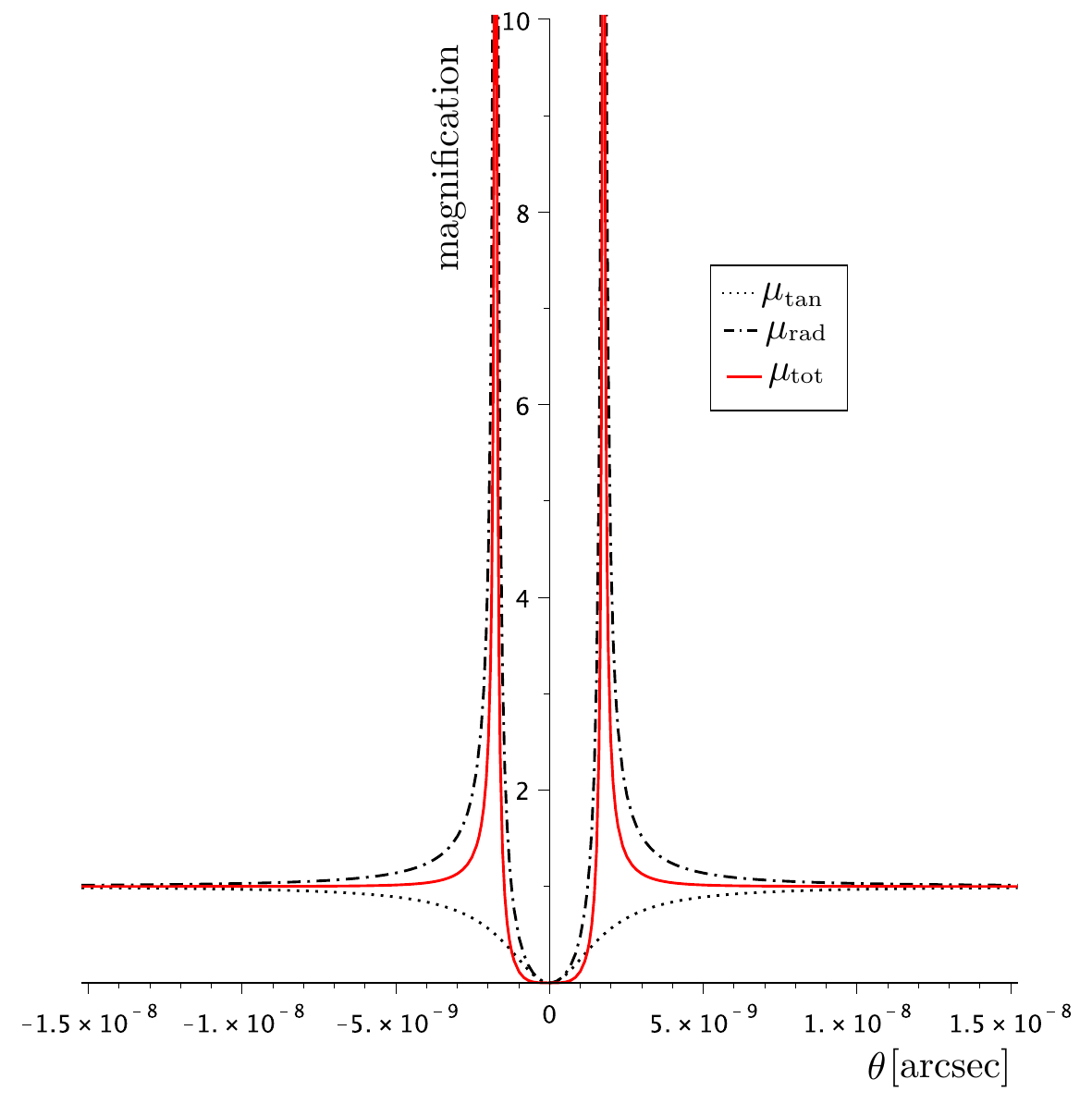}
    
    (d) $\ell = -1$\\
    \quad {\small\textit{(Theoretically Acceptable Value)}}
  \end{minipage}
};

\end{tikzpicture}

\vspace{0.2cm}
\caption{The magnifications—tangential $\mu_{\rm tan}$ (dotted lines), radial $\mu_{\rm rad}$ (dash-dotted lines), and total $\mu$ (continuous curves)—are plotted as functions of the image position $\theta$ for four SKRCS BH cases with $\beta=10$. Panels (a)–(d) correspond to different values of $\ell$ as labeled. The singularities of $\mu_{\rm tan}$ and $\mu_{\rm rad}$ give the positions of the tangential and radial critical curves, respectively. Here $|M|=1\, M_{\odot}$, $D_{\rm s}=0.05\text{ Mpc}$ and $D_{\rm l}=0.01\text{ Mpc}$. Angles are in arcseconds: $1\text{ arcsec}=4.848\times10^{-6}\text{ rad}$.}
\label{isfigLENSBETA10}
\end{figure}

Figure~\ref{isfigLENSBETA10} represents the extreme CS parameter regime ($\beta=10$) where topological defect contributions systematically suppress LV effects, demonstrating the complex interplay between these two fundamental modifications to gravitational lensing phenomena. The observationally constrained scenarios in panels (a) and (b) exhibit dramatic suppression of magnification amplitudes compared to previous configurations, with critical curve positioning approaching conventional Schwarzschild behavior despite substantial underlying spacetime modifications. These results suggest that strong CS configurations may effectively mask LV signatures in observational surveys, potentially explaining the absence of detected LV effects in certain astrophysical environments. The theoretical exploration regimes in panels (c) and (d) reveal systematic asymptotic behavior where magnification characteristics approach limiting values determined primarily by CS geometry rather than LV contributions, indicating hierarchical parameter dominance relationships within SKRCS configurations. The striking convergence of magnification profiles across different LV parameter values in this regime suggests that strong CS fields establish dominant geometric constraints that overwhelm subtle LV modifications, providing crucial theoretical insights for interpreting null results in observational searches for exotic gravitational physics.

The systematic exploration of magnification characteristics across the complete SKRCS parameter space reveals fundamental scaling relationships between LV and CS contributions to gravitational lensing phenomena. The comprehensive visualization demonstrates that parameter variations induce systematic modifications in both critical curve positioning and magnification amplitude scaling, establishing quantitative frameworks for distinguishing SKRCS configurations from conventional Schwarzschild spacetimes through precision astronomical observations. The comparative analysis reveals significant structural differences contingent upon the specific values of $(\ell,\beta)$ parameter combinations. In the Schwarzschild limit $(\ell=0,\beta=1)$, magnification behavior conforms to established patterns documented in classical lensing theory. However, systematic deviations from this baseline demonstrate progressively increasing modifications as parameters deviate from standard values, indicating nonlinear coupling effects between LV and CS contributions that could potentially amplify observable signatures beyond simple additive contributions. Moreover, the theoretical framework established through this magnification analysis provides quantitative predictions for precision gravitational lensing observations that could potentially detect and characterize SKRCS spacetime modifications. The systematic parameter dependencies demonstrated in the magnification analysis suggest optimal observational strategies for constraining exotic physics contributions, particularly through statistical analysis of large gravitational lensing surveys where subtle systematic effects might be detectable despite individual measurement uncertainties \cite{is34,is35}.

\section{Photon Lights in Schwarzschild-like BH in KR gravity with C\lowercase{o}S}\label{isec4}

The theoretical exploration of gravitational phenomena in modified spacetime geometries necessitates comprehensive investigation of alternative topological configurations beyond conventional CS-pierced spacetimes. The incorporation of CoS distributions into KR gravity frameworks represents a fundamental advancement in understanding exotic gravitational physics, establishing novel theoretical paradigms for investigating LV mechanisms within cosmologically motivated string theory contexts \cite{is08,sec2is37}.

The comprehensive theoretical framework for KR gravity coupled to CoS configurations emerges from the Einstein-Hilbert action extended through nonminimal coupling to self-interacting KR fields \cite{is16,is08}:

\begin{eqnarray}\label{AT}
\mathcal{S}= \frac{1}{\kappa}\int \ d^{4}x \sqrt{-g}\bigg[R-2\Lambda - \frac{1}{6} H^{\mu \nu \rho} H_{\mu \nu \rho}-V(\mathcal{B}^{\mu \nu}\mathcal{B}_{\mu \nu})+\xi_2 \mathcal{B}^{\rho \mu } \mathcal{B}_{\mu}^\nu R_{\rho \nu} + \xi_3 \mathcal{B}^{\mu \nu } \mathcal{B}_{\mu \nu} R \bigg] +\int \ d^{4}x \sqrt{-g}\mathcal\,{L}_{M}, 
\end{eqnarray}

where $\kappa = 8\pi G$ represents the gravitational coupling constant, $\xi_2$ and $\xi_3$ denote the coupling constants governing the interaction between gravitational and KR field components, $H_{\mu \nu \rho} = \partial_{[\mu} \mathcal{B}_{\nu \rho]}$ characterizes the KR field strength tensor, and $\Lambda$ represents the cosmological constant. The self-interaction potential $V(\mathcal{B}^{\mu \nu}\mathcal{B}_{\mu \nu})$ maintains theoretical invariance under local Lorentz transformations while enabling spontaneous symmetry breaking mechanisms.

The action explicitly incorporates gauge symmetry breaking through nonminimal coupling terms and the smooth potential $V(\mathcal{B}^{\mu \nu}\mathcal{B}_{\mu \nu}\pm b^2)$, inducing spontaneous LV through the mechanism of dynamical symmetry breaking. This fundamental mechanism generates a non-trivial VEV for the KR field configuration: $\langle \mathcal{B}^{\mu \nu}\rangle = b^{\mu \nu}$, establishing the theoretical foundation for systematic LV effects in gravitational phenomena.

Variational principles applied to the action (\ref{AT}) with respect to the metric tensor $g_{\mu \nu}$ yield the modified Einstein field equations:
\begin{equation}
R_{\mu \nu }-\frac{1}{2}g_{\mu \nu }R+\Lambda g_{\mu \nu }=T_{\mu \nu }^{KR}+T_{\mu \nu }^{M},  \label{eom11}
\end{equation}

where $T_{\mu \nu }^{KR}$ and $T_{\mu \nu }^{M}$ represent the energy-momentum tensors for KR and conventional matter fields, respectively.

The theoretical incorporation of CoS distributions requires systematic formulation of the corresponding energy-momentum tensor through variational principles applied to the string action. The CoS energy-momentum tensor assumes the form:
\begin{equation}
   T_{\mu\nu}^{CoS}=2 \frac{\partial}{\partial g_{\mu \nu}}\mathcal{M}\sqrt{-\frac{1}{2}\Sigma^{\mu \nu}\,\Sigma_{\mu\nu}} =\frac{\rho \,\Sigma_{\alpha\nu}\, \,\Sigma_{\mu}^\alpha }{\sqrt{-\gamma}}, 
\end{equation}

where $\rho$ denotes the proper density distribution of the CoS configuration, and $\Sigma^{\mu \nu}$ represents the string world-sheet tensor characterizing the geometric properties of the string cloud distribution.

Application of conservation laws through the covariant derivative $\nabla_\mu T_{\mu\nu}^{CoS}=0$ yields the non-vanishing energy-momentum tensor components:
\begin{equation}
 T_{t}^{CoS t}=T_{r}^{CoS r}=-\frac{\alpha}{r^2},
\end{equation}  

where $\alpha$ represents the fundamental CoS parameter characterizing the strength and distribution of the string cloud configuration within the spacetime geometry.

The synthesis of KR gravity modifications with CoS topological configurations yields the SKRCoS BH spacetime, representing a novel class of solutions that systematically incorporates both LV mechanisms and distributed string matter sources \cite{is19}. The SKRCoS line element is characterized by:
\begin{eqnarray}
   ds^2=-\mathcal{B}(r,\ell)\,dt^2+\frac{dr^2}{\mathcal{B}(r,\ell)}+r^2\,(d\theta^2+\sin^2 \theta\,d\phi^2),\label{d1}
\end{eqnarray}

where the modified metric function incorporates both LV and CoS contributions:
\begin{equation}
    \mathcal{B}(r,\ell)=\frac{1}{1-\ell}-\alpha-\frac{2\,M}{r}=\eta-\frac{2\,M}{r},\label{d2}
\end{equation}

The composite parameter $\eta = \frac{1}{1-\ell} - \alpha$ systematically encodes both LV effects through $\ell$ and CoS contributions through $\alpha$, with the constraint $0 < \alpha < 1$ ensuring physical consistency of the CoS distribution \cite{is18}.

The theoretical framework exhibits elegant limiting behavior: $\alpha=0$ recovers the pure KR BH solution \cite{is09}, while $\ell=0$ reduces to the conventional Schwarzschild BH with CoS configuration \cite{is19}. The simultaneous limits $\ell=0$ and $\alpha=0$ reproduce standard Schwarzschild spacetime.

The investigation of photon trajectories in SKRCoS geometries requires systematic analysis of null geodesics restricted to the equatorial plane $\theta=\pi/2$ for mathematical tractability while preserving the essential physics of gravitational lensing phenomena. The Lagrangian density function for particle motion in SKRCoS spacetime becomes:
\begin{equation}
\mathcal{L}=\frac{1}{2}\,\Bigg[-\mathcal{B}\,\left(\frac{dt}{d\tau}\right)^2+\frac{1}{\mathcal{B}}\,\left(\frac{dr}{d\tau}\right)^2+r^2\,\left(\frac{d\phi}{d\tau}\right)^2\Bigg].\label{d3}
\end{equation}

Application of Euler-Lagrange variational principles yields the fundamental geodesic equations:
\begin{eqnarray}
    &&\frac{dt}{d\tau}=\frac{\mathrm{E}}{\mathcal{B}}\,,\nonumber\\
    &&\frac{d\phi}{d\tau}=\frac{\mathrm{L}}{r^2}.\label{d4}
\end{eqnarray}

The radial equation for null geodesics assumes the canonical form:
\begin{equation}
    \left(\frac{dr}{d\tau}\right)^2+V_\text{eff}(r)=\mathrm{E}^2,\label{d5}
\end{equation}

where the effective potential systematically incorporates both LV and CoS modifications:
\begin{equation}
    V_\text{eff}(r)=\frac{\mathrm{L}^2}{r^2}\,\left(\eta-\frac{2\,M}{r}\right)=\frac{\mathrm{L}^2}{r^2}\,\left(\frac{1}{1-\ell}-\alpha-\frac{2\,M}{r}\right).\label{d6}
\end{equation}

Combining the geodesic equations (\ref{d4}) and (\ref{d5}), the fundamental radial orbit equation becomes:
\begin{equation}
    \left(\frac{dr}{d\phi}\right)^2=r^4\,\left[\frac{1}{\gamma^2}-\frac{1}{r^2}\,\left(\eta-\frac{2\,M}{r}\right)\right].\label{d7}
\end{equation}

The determination of photon deflection angles requires integration of the orbit equation over the complete photon trajectory from asymptotic infinity to the turning point and back to asymptotic infinity. The differential form of the orbit equation is:
\begin{equation}
    d\phi=\frac{dr}{r^2\,\left[\frac{1}{\gamma^2}-\frac{1}{r^2}\,\left(\eta-\frac{2\,M}{r} \right)\right]^{1/2}}.\label{d8}
\end{equation}

The total deflection angle for SKRCoS photon trajectories is:
\begin{equation}
    \hat{\alpha}_\text{II}=\Delta\phi-\pi,\label{d9}
\end{equation}

where the total angular displacement is:
\begin{equation}
    \Delta\phi=2\,\int^{\infty}_{r_0}\,\frac{dr}{r^2\,\left[\frac{1}{\gamma^2}-\frac{1}{r^2}\,\left(\eta-\frac{2\,M}{r} \right)\right]^{1/2}}.\label{d10}
\end{equation}

The turning point $r_0$ is determined by the condition $\frac{dr}{d\tau}=0$, yielding:
\begin{equation}
\frac{1}{\gamma^2}=\frac{1}{r^2_0}\,\left(\eta-\frac{2\,M}{r_0} \right).\label{d11}     
\end{equation}

Through the canonical transformation $r=\frac{1}{v}$, the deflection integral becomes:
\begin{equation}
    \Delta\phi=2\,\int^{v_0}_{0}\,\frac{dv}{\sqrt{\eta\,(v_0^2-v^2)-2\,M\,(v_0^3-v^3)}}.\label{d12}
\end{equation}

The integrand factorizes into cubic polynomial form:
\begin{equation}
    \eta\,(v_0^2-v^2)-2\,M\,(v_0^3-v^3)=a_1\,v^3-b_1\,v^2-c_1=a_1\,(v-v_1)\,(v-v_2)\,(v-v_3),\label{d13}
\end{equation}

with coefficients:
\begin{equation}
    a_1=2\,M,\quad\quad b_1=\eta,\quad\quad c_1=2\,M\,v^3_0-v^2_0\,\eta.\label{d14}
\end{equation}

\subsection{Deflection Angle via Perturbation Method}

The perturbative analysis of SKRCoS deflection angles employs systematic expansion techniques to obtain approximate analytical solutions with controlled accuracy. Starting with the inverse radial coordinate transformation $r = 1/u(\phi)$, the fundamental trajectory equation becomes:
\begin{equation}
\left( \frac{du}{d\phi} \right)^2 = \eta u^2 - 2M u^3 - \frac{1}{\gamma^2}, \label{per1}
\end{equation}

where, as previously stated, $\eta = \frac{1}{1 - \ell} - \alpha$ systematically encodes both LV and CoS modifications, and $\gamma = L/E$ represents the impact parameter.

Differentiation with respect to $\phi$ yields the second-order differential equation:
\begin{equation}
\frac{d^2 u}{d\phi^2} + \eta u = 3Mu^2. \label{per2}
\end{equation}

The perturbative solution employs the expansion:
\begin{equation}
u = u_0 + \varepsilon u_1 + \varepsilon^2 u_2 + \cdots, \label{per3}
\end{equation}

where $\varepsilon = M/\gamma$ represents the gravitational strength parameter. The homogeneous equation reads
\begin{equation}
\frac{d^2 u_0}{d\phi^2} + \eta u_0 = 0, \label{per4}
\end{equation}

yields the solution:
\begin{equation}
u_0 = u_\perp \cos(\sqrt{\eta} \phi), \label{per5}
\end{equation}

where $u_\perp = 1/\gamma$ corresponds to the inverse closest approach distance. Hence, we get the following inhomogeneous equation
\begin{equation}
\frac{d^2 u_1}{d\phi^2} + \eta u_1 = 3M u_\perp^2 \cos^2(\sqrt{\eta} \phi), \label{per6}
\end{equation}

admits the analytical solution:
\begin{equation}
u_1 = \frac{u_\perp^2 M}{2\eta} \left[3 - \cos(2\sqrt{\eta} \phi)\right]. \label{per7}
\end{equation}

Equation of the second-order correction becomes
\begin{equation}
\frac{d^2 u_2}{d\phi^2} + \eta u_2 = 6M u_\perp \cos(\sqrt{\eta} \phi) u_1, \label{per8}
\end{equation}

yields:
\begin{equation}
u_2 = \frac{3M^2 u_\perp^3}{16\eta^{5/2}} \left[ \cos(3\sqrt{\eta} \phi)\sqrt{\eta} + 20 \phi \sin(\sqrt{\eta} \phi) \right]. \label{per9}
\end{equation}

The full trajectory becomes:
\begin{align}
u(\phi) = u_\perp \Bigg[ &\cos(\sqrt{\eta} \phi) + \frac{\varepsilon M u_\perp}{2\eta} \left(3 - \cos(2\sqrt{\eta} \phi)\right)  + \frac{3\varepsilon^2 M^2 u_\perp^2}{16\eta^{5/2}} \left(\cos(3\sqrt{\eta} \phi) \sqrt{\eta} + 20 \phi \sin(\sqrt{\eta} \phi)\right) \Bigg]. \label{per10}
\end{align}

The deflection angle determination exploits symmetry through:
\begin{equation}
\phi = \frac{\pi}{2} + \psi \quad \Rightarrow \quad \hat{\alpha}_{\text{pert}} = 2\psi. \label{per11}
\end{equation}

Expanding to second order in $\varepsilon = M/\gamma$:
\begin{equation}
\hat{\alpha}_{\text{pert}} \approx \frac{4M}{\gamma \eta} + \frac{15\pi M^2}{4\gamma^2 \eta^2}. \label{per12}
\end{equation}

Expressing in terms of closest approach distance $r_c$:
\begin{equation}
\frac{1}{r_c} = u_\perp + M u_\perp^2 + \frac{3}{16} M^2 u_\perp^3, \label{per13}
\end{equation}

yields:
\begin{equation}
\hat{\alpha}_{\text{pert}} \approx \frac{4M}{\eta r_c} + \frac{M^2}{\eta^2 r_c^2} \left( \frac{15\pi}{4} - 4 \right). \label{per14}
\end{equation}

This expression explicitly demonstrates the systematic contributions of LV parameter $\ell$ and CoS parameter $\alpha$ through $\eta$.

\subsection{Analytical Deflection Angle via Elliptic Integrals}

The exact analytical determination of SKRCoS deflection angles employs elliptic integral methodologies to obtain precise mathematical expressions valid across the complete parameter space. The radial equation in inverse coordinates $v = 1/r$ assumes the cubic polynomial form:
\begin{equation}
\left( \frac{dv}{d\phi} \right)^2 = 2M \left(v - v_1\right)\left(v_2 - v\right)\left(v_3 - v\right),
\end{equation}

where $v_1 < 0 < v_2 = v_0 < v_3$ represent the roots of the underlying cubic polynomial.

The periastron parameter formulation introduces $P = 1/v_2$ and auxiliary parameter:
\begin{equation}
R = P\eta,
\end{equation}

enabling explicit root expressions:
\begin{align}
v_1 &= \frac{R - 2M - Q}{4MP}, \\
v_2 &= \frac{1}{P}, \\
v_3 &= \frac{R - 2M + Q}{4MP},
\end{align}

with discriminant:
\begin{equation}
Q^2 = (R - 2M)(R + 6M).
\end{equation}

The impact parameter becomes:
\begin{equation}
\gamma = \frac{\eta^{3/2} P^{3/2}}{\sqrt{P\eta - 2M}}.
\end{equation}

The exact deflection angle integral:
\begin{equation}
\hat{\alpha}_{II} = 2 \int_0^{v_2} \frac{dv}{\sqrt{2M(v - v_1)(v_2 - v)(v_3 - v)}} - \pi,
\end{equation}

admits elliptic integral representation:
\begin{equation}
\hat{\alpha}_{II} = \sqrt{\frac{2}{M}} \left[ \frac{2 F(\Psi_1, k)}{\sqrt{v_3 - v_1}} - \frac{2 F(\Psi_2, k)}{\sqrt{v_3 - v_1}} \right] - \pi,
\end{equation}

where the elliptic integral parameters are:
\begin{align}
k^2 &= \frac{v_2 - v_1}{v_3 - v_1}, \\
\Psi_1 &= \frac{\pi}{2}, \\
\Psi_2 &= \sin^{-1} \sqrt{ \frac{-v_1}{v_2 - v_1} } = \sin^{-1} \sqrt{\frac{Q + 2M - R}{Q + 6M - R}}.
\end{align}

This exact analytical framework captures the complete gravitational deflection physics in SKRCoS spacetimes, systematically incorporating both LV and CoS contributions through the parameter dependencies in $\ell$ and $\alpha$. The classical Schwarzschild limit emerges when $\ell = 0$ and $\alpha = 0$, confirming the mathematical consistency of the theoretical formulation.

The comprehensive analytical treatment of SKRCoS deflection phenomena establishes fundamental theoretical frameworks for understanding the interplay between LV mechanisms and distributed string matter sources in gravitational lensing contexts. These theoretical developments contribute to the broader understanding of modified gravity phenomena while providing solid pathways for experimental verification through precision gravitational lensing observations \cite{sec2is42,sec2is43}.

\section{Magnification of SKRCoS BHs} \label{isec5}

The gravitational lensing phenomenon represents one of the most profound manifestations of exotic spacetime curvature, particularly when investigating BH configurations that incorporate both LV mechanisms and CoS topological structures. The magnification effects induced by SKRCoS BHs establish unprecedented theoretical frameworks for probing fundamental spacetime deformations characterized by the interplay between the LV parameter $\ell$ and the CoS tension parameter $\alpha$. This comprehensive investigation systematically extends conventional Schwarzschild lensing paradigms to encompass these exotic modifications, analyzing how photon ray magnification undergoes systematic alterations due to the combined influence of LV and CoS deformation parameters \cite{is20,is21}.

The spherical symmetry inherent in SKRCoS geometries ensures that gravitational lensing affects exclusively radial distance measurements while preserving azimuthal angle invariance, yielding the fundamental relationship between image position $\Gamma_{\rm CoS}$ and source position $\tilde{\theta}$ \cite{is27}:

\begin{equation}
\Gamma_{\rm CoS} = \tilde{\theta} + \frac{\bar{\theta}_{\xi}^2}{\tilde{\theta}},
\end{equation}

where the modified Einstein angle systematically incorporates both LV and CoS contributions:

\begin{equation}
\bar{\theta}_{\xi}^2 = \theta_{\xi}^2 \left( \frac{1}{\eta} + \frac{15 \pi G M}{16 \tilde{\theta} c^2 \eta^2 D_l} \right),
\end{equation}

with the standard Einstein angle defined as:
\begin{equation}
\tilde{\theta}_{\xi}^2 = \frac{4 G M}{c^2} \times \frac{D_{\mathrm{ls}}}{D_{\mathrm{s}} D_l}.
\end{equation}

The quantitative determination of magnification effects requires sophisticated analysis of how gravitational lensing modifies the cross-sectional area of photon ray bundles, directly influencing the observed brightness characteristics of lensed astronomical sources. For infinitesimally small source configurations, the total magnification factor $\tilde{\mu}_{\rm tot}$ is expressed through the Jacobian determinant of the lens mapping \cite{sec2is47}:

\begin{equation}
\mu^{-1} = \left| \frac{\Gamma_{\rm CoS}}{\tilde{\theta}} \frac{d\Gamma_{\rm CoS}}{d\tilde{\theta}} \right|.
\end{equation}

This theoretical formulation demonstrates that lensed images undergo systematic magnification or demagnification by the factor $|\mu|$, with the magnification characteristics directly correlated to the spacetime curvature modifications introduced by LV and CoS parameters in the near-BH regime. As being stated before, the systematic decomposition of magnification effects into tangential and radial components provides fundamental insights into the geometric structure of SKRCoS gravitational lensing phenomena. These components exhibit distinct contributions to overall magnification, characterized by divergent behavior at critical curves where magnification theoretically approaches infinity \cite{is28}.

The tangential magnification component is expressed as:

\begin{equation}
\mu_{\rm tan} = \left| \frac{\Gamma_{\rm CoS}}{\tilde{\theta}} \right|^{-1} = \frac{\tilde{\theta}^2}{\tilde{\theta}^2 + \bar{\theta}_{\xi}^2},
\end{equation}

while the radial magnification exhibits:

\begin{equation}
\mu_{\rm rad} = \left| \frac{d\Gamma_{\rm CoS}}{d\tilde{\theta}} \right|^{-1} = \frac{\tilde{\theta}^2}{\tilde{\theta}^2 - \bar{\theta}_{\xi}^2}.
\end{equation}

A distinctive characteristic emerges through this analysis: $\mu_{\rm rad}$ exhibits mathematical singularities at $\tilde{\theta} = \tilde{\theta}_{\rm E}$, corresponding to the angular radius of radial critical curves, while $\mu_{\rm tan}$ maintains finite values throughout the accessible parameter space.

As demonstrated in the deflection angle analysis of Eq.~\eqref{per12}, photon trajectories passing in close proximity to the BH experience substantial modifications through these exotic parameters. These alterations become particularly pronounced for light rays traversing the strong gravitational field regime, where higher-order terms in the deflection angle expansion contribute significantly to total magnification characteristics. The theoretical incorporation of both the LV parameter $\ell$ and the CoS parameter $\alpha$ generates remarkable structural modifications in the lensing curve topology, establishing distinctive observational signatures that potentially enable discrimination between SKRCoS BH configurations and alternative compact object models. The magnification patterns predicted through this theoretical framework establish quantitative benchmarks for future astronomical observations of gravitationally lensed systems, offering novel experimental pathways for constraining the fundamental nature of exotic BH physics beyond conventional GR predictions \cite{is30}.

\begin{figure}[H]
\centering
\begin{tikzpicture}[inner sep=0pt, outer sep=0pt]

\draw[thick] (0.5\linewidth,0) -- (0.5\linewidth,-0.9\linewidth);
\draw[thick] (0,-0.45\linewidth) -- (\linewidth,-0.45\linewidth);

\node[anchor=north west] at (0,0){
  \begin{minipage}[t][0.45\linewidth][c]{0.45\linewidth}
    \centering
    \includegraphics[width=0.8\linewidth]{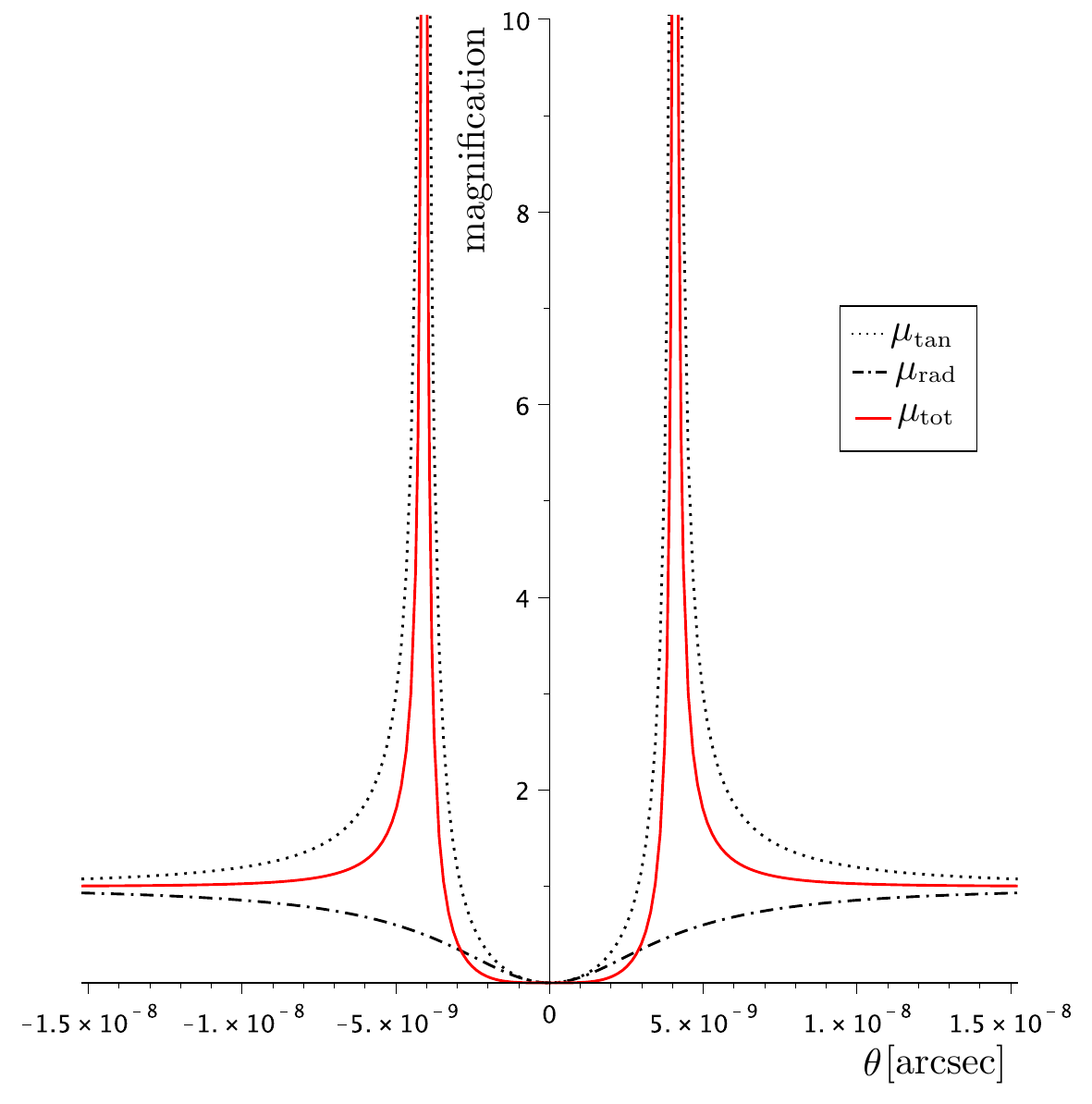}
    
    (a) $\ell = 0.06092$ \\
    \quad {\small\textit{(Within Milky Way Galaxy Bound: $-0.18502<\ell<0.06093$)}}
  \end{minipage}
};

\node[anchor=north west] at (0.5\linewidth,0){
  \begin{minipage}[t][0.45\linewidth][c]{0.45\linewidth}
    \centering
    \includegraphics[width=0.8\linewidth]{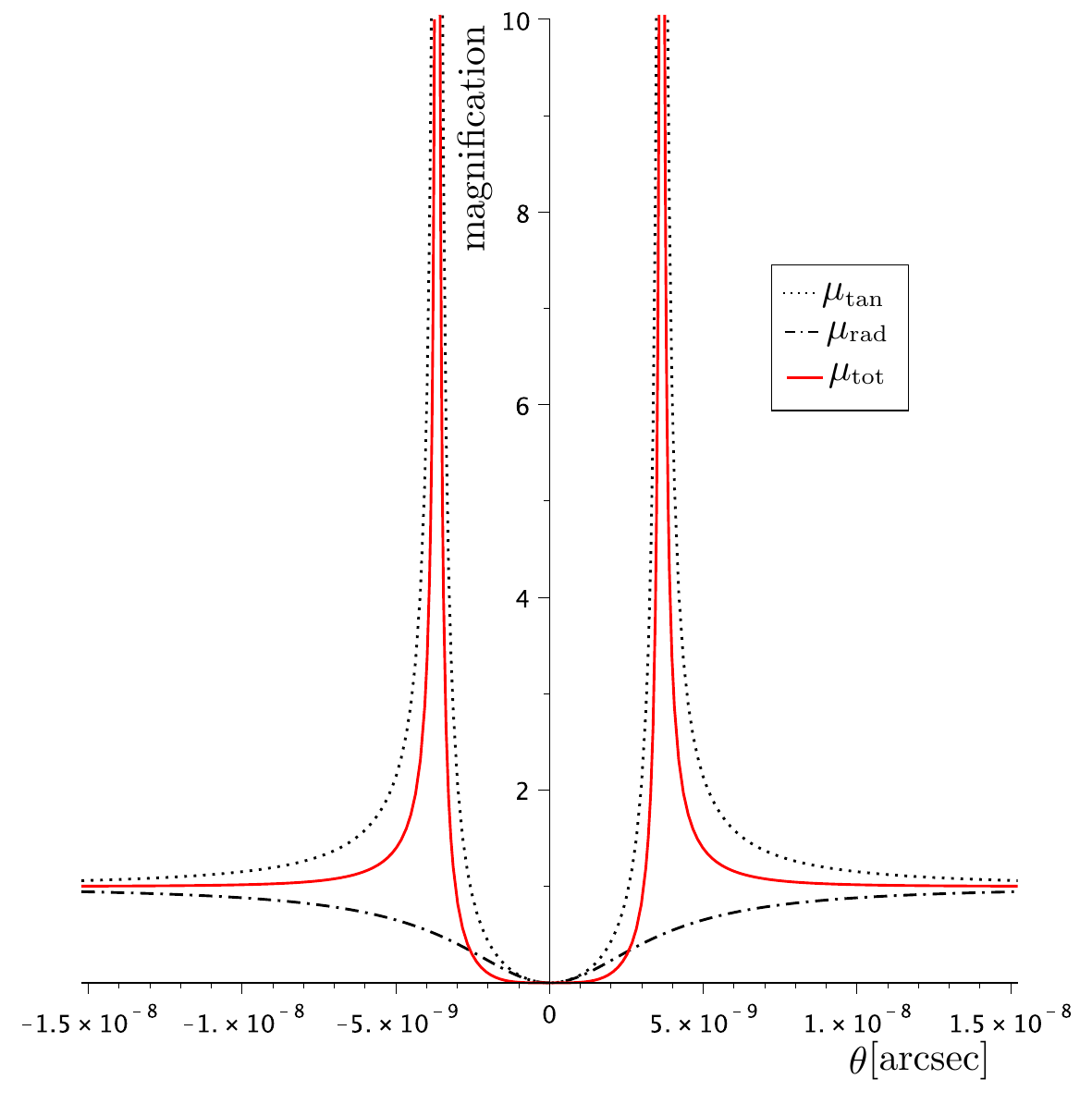}
    
    (b) $\ell = -0.18501$ \\
    \quad {\small\textit{(Within Milky Way Galaxy Bound: $-0.18502<\ell<0.06093$)}}
  \end{minipage}
};

\node[anchor=north west] at (0,-0.45\linewidth){
  \begin{minipage}[t][0.45\linewidth][c]{0.45\linewidth}
    \centering
    \includegraphics[width=0.8\linewidth]{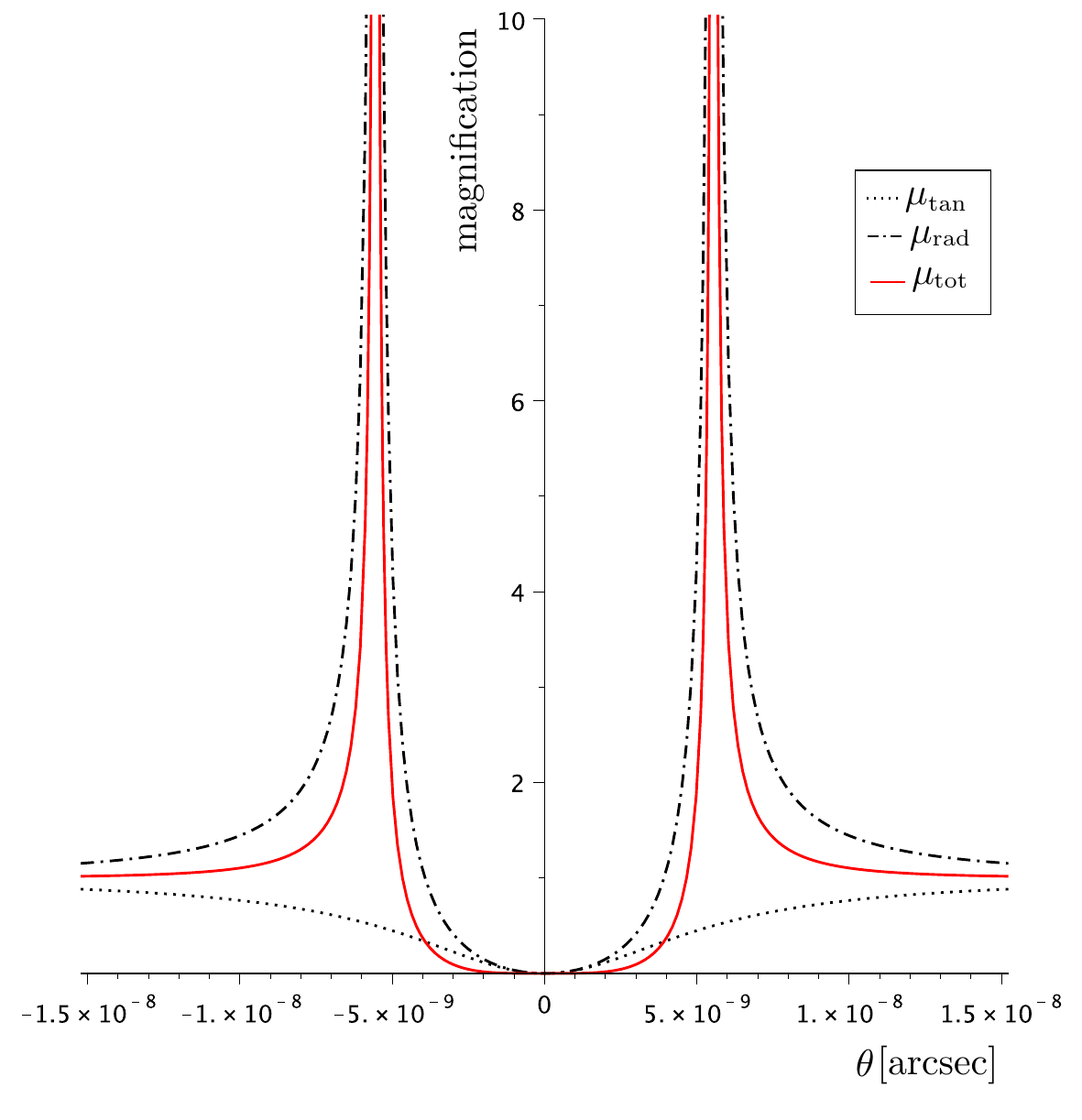}
    
    (c) $\ell = 0.6$\\
    \quad {\small\textit{(Theoretically Acceptable Value)}}
  \end{minipage}
};

\node[anchor=north west] at (0.5\linewidth,-0.45\linewidth){
  \begin{minipage}[t][0.45\linewidth][c]{0.5\linewidth}
    \centering
    \includegraphics[width=0.8\linewidth]{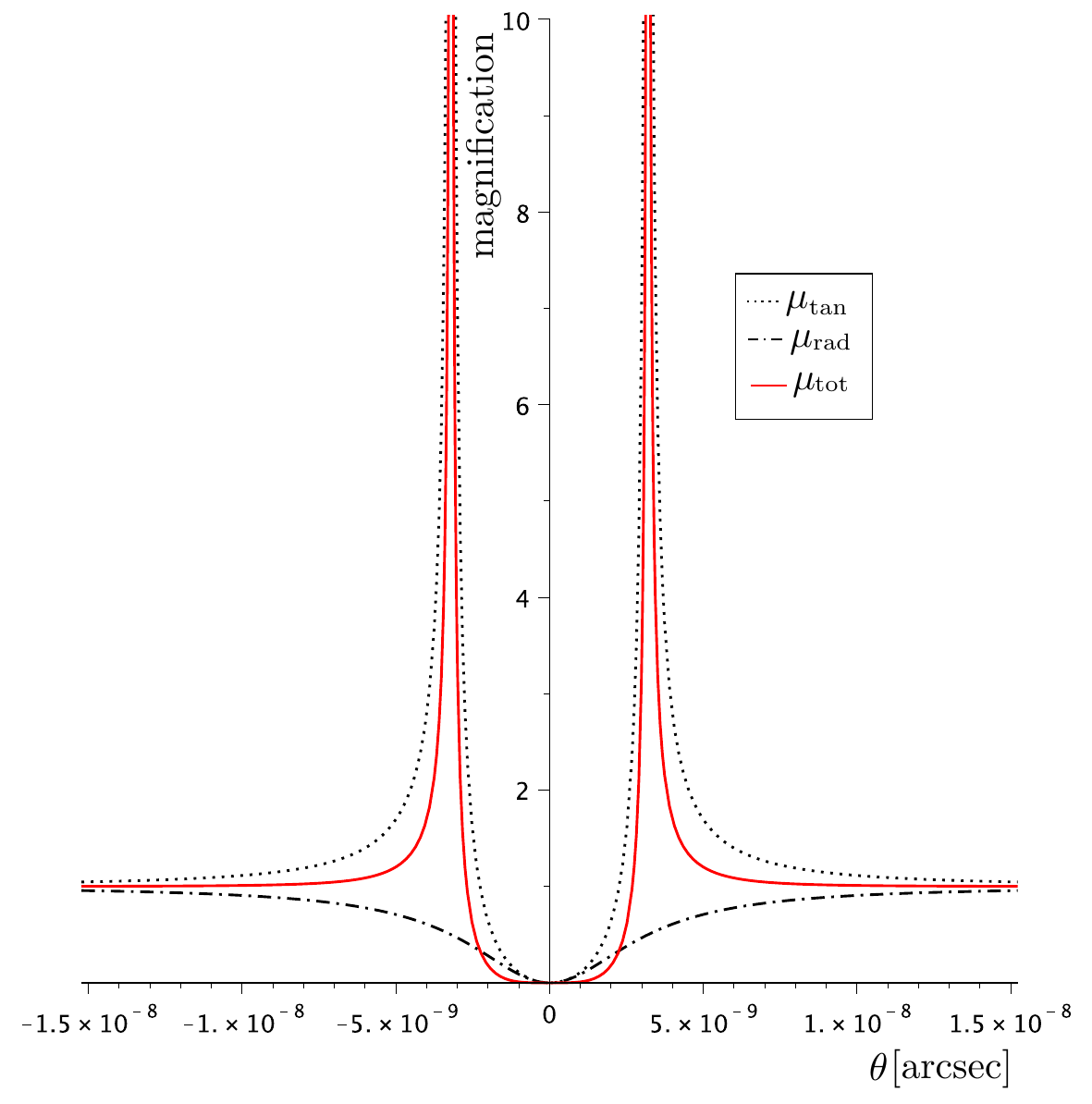}
    
    (d) $\ell = -1$\\
    \quad {\small\textit{(Theoretically Acceptable Value)}}
  \end{minipage}
};

\end{tikzpicture}

\vspace{0.2cm}
\caption{The magnifications—tangential $\mu_{\rm tan}$ (dotted lines), radial $\mu_{\rm rad}$ (dash-dotted lines), and total $\mu$ (continuous curves)—are plotted as functions of the image position $\theta$ for four SKRCoS BH cases with $\alpha=0.01$. Panels (a)–(d) correspond to different values of $\ell$ as labeled. The singularities of $\mu_{\rm tan}$ and $\mu_{\rm rad}$ give the positions of the tangential and radial critical curves, respectively. Here $|M|=1\, M_{\odot}$, $D_{\rm s}=0.05\text{ Mpc}$ and $D_{\rm l}=0.01\text{ Mpc}$. Angles are in arcseconds: $1\text{ arcsec}=4.848\times10^{-6}\text{ rad}$.}
\label{isfigLENSalpha001}
\end{figure}

Figure~\ref{isfigLENSalpha001} demonstrates the systematic influence of LV parameter variations within the minimal CoS regime ($\alpha=0.01$), revealing fundamental modifications to conventional gravitational lensing behavior under conditions where CoS contributions remain negligible compared to LV effects. The observationally constrained scenarios in panels (a) and (b) exhibit relatively modest deviations from standard Schwarzschild magnification characteristics, with critical curve positioning shifts that remain within potentially detectable ranges using contemporary precision astrometric techniques. The theoretical exploration regimes in panels (c) and (d) reveal progressively dramatic magnification modifications that, while exceeding current observational bounds, establish crucial theoretical benchmarks for understanding the fundamental scaling relationships between LV contributions and observable lensing signatures. The systematic comparison between positive and negative LV parameter configurations demonstrates pronounced asymmetries in magnification response characteristics, suggesting that gravitational lensing observations could potentially discriminate between different theoretical implementations of LV mechanisms through statistical analysis of large astronomical surveys. The minimal CoS contribution in this configuration enables clear isolation of pure LV effects, providing essential theoretical foundations for disentangling the relative contributions of fundamental spacetime modifications versus topological defect structures in future observational programs.


\begin{figure}[ht!]
\centering
\begin{tikzpicture}[inner sep=0pt, outer sep=0pt]

\draw[thick] (0.5\linewidth,0) -- (0.5\linewidth,-0.9\linewidth);
\draw[thick] (0,-0.45\linewidth) -- (\linewidth,-0.45\linewidth);

\node[anchor=north west] at (0,0){
  \begin{minipage}[t][0.45\linewidth][c]{0.45\linewidth}
    \centering
    \includegraphics[width=0.8\linewidth]{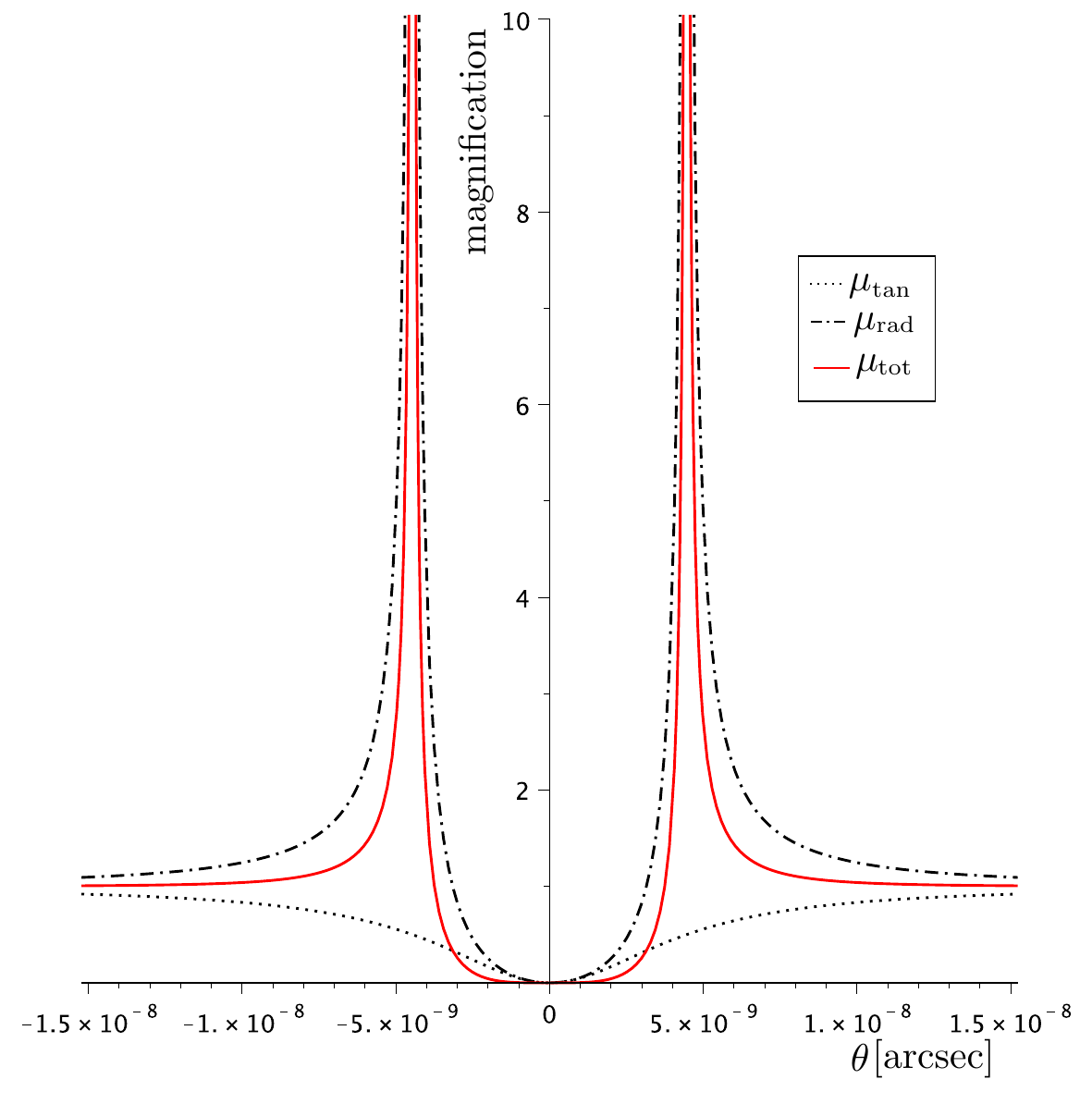}
    
    (a) $\ell = 0.06092$ \\
    \quad {\small\textit{(Within Milky Way Galaxy Bound: $-0.18502<\ell<0.06093$)}}
  \end{minipage}
};

\node[anchor=north west] at (0.5\linewidth,0){
  \begin{minipage}[t][0.45\linewidth][c]{0.45\linewidth}
    \centering
    \includegraphics[width=0.8\linewidth]{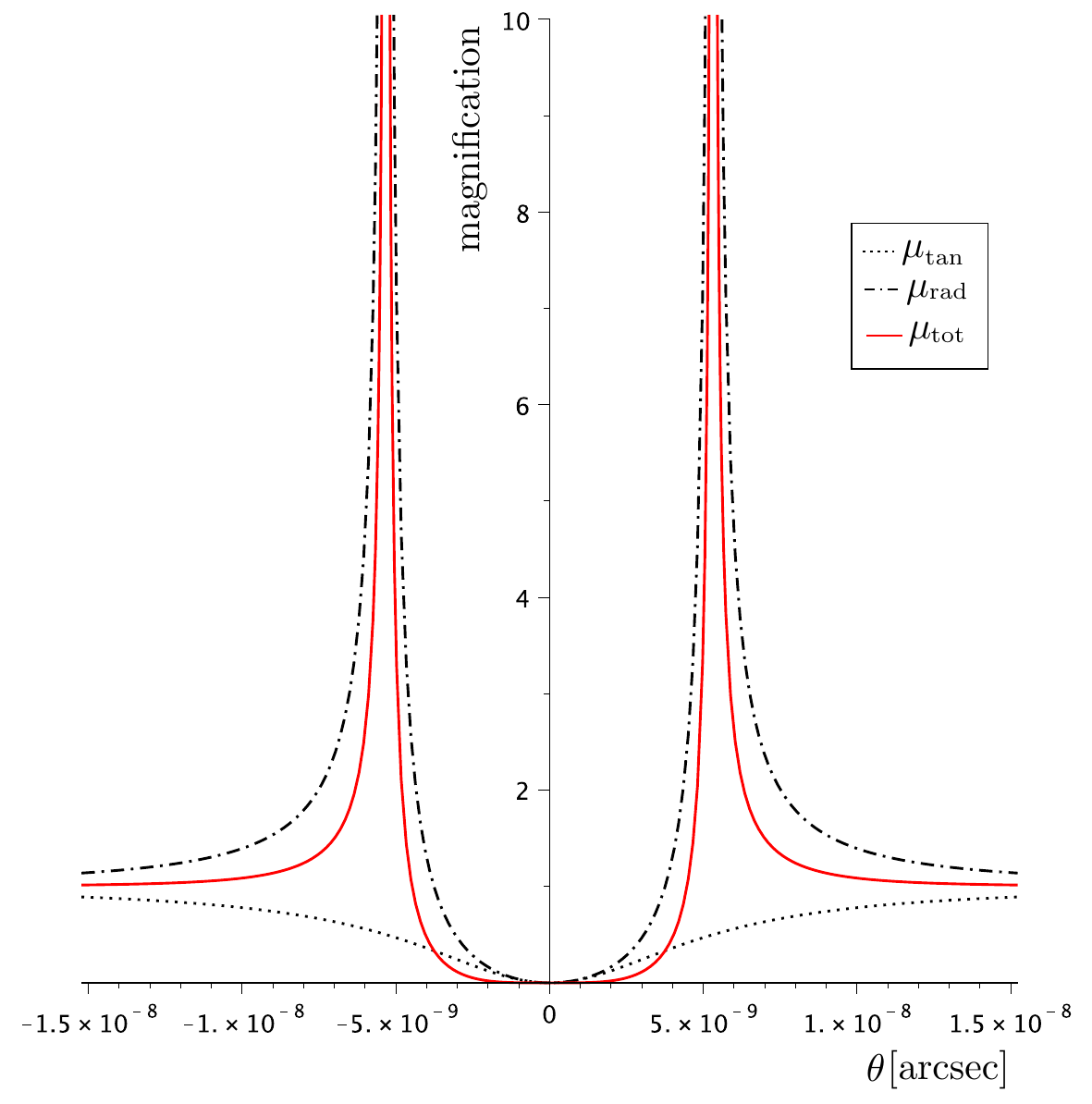}
    
    (b) $\ell = -0.18501$ \\
    \quad {\small\textit{(Within Milky Way Galaxy Bound: $-0.18502<\ell<0.06093$)}}
  \end{minipage}
};

\node[anchor=north west] at (0,-0.45\linewidth){
  \begin{minipage}[t][0.45\linewidth][c]{0.45\linewidth}
    \centering
    \includegraphics[width=0.8\linewidth]{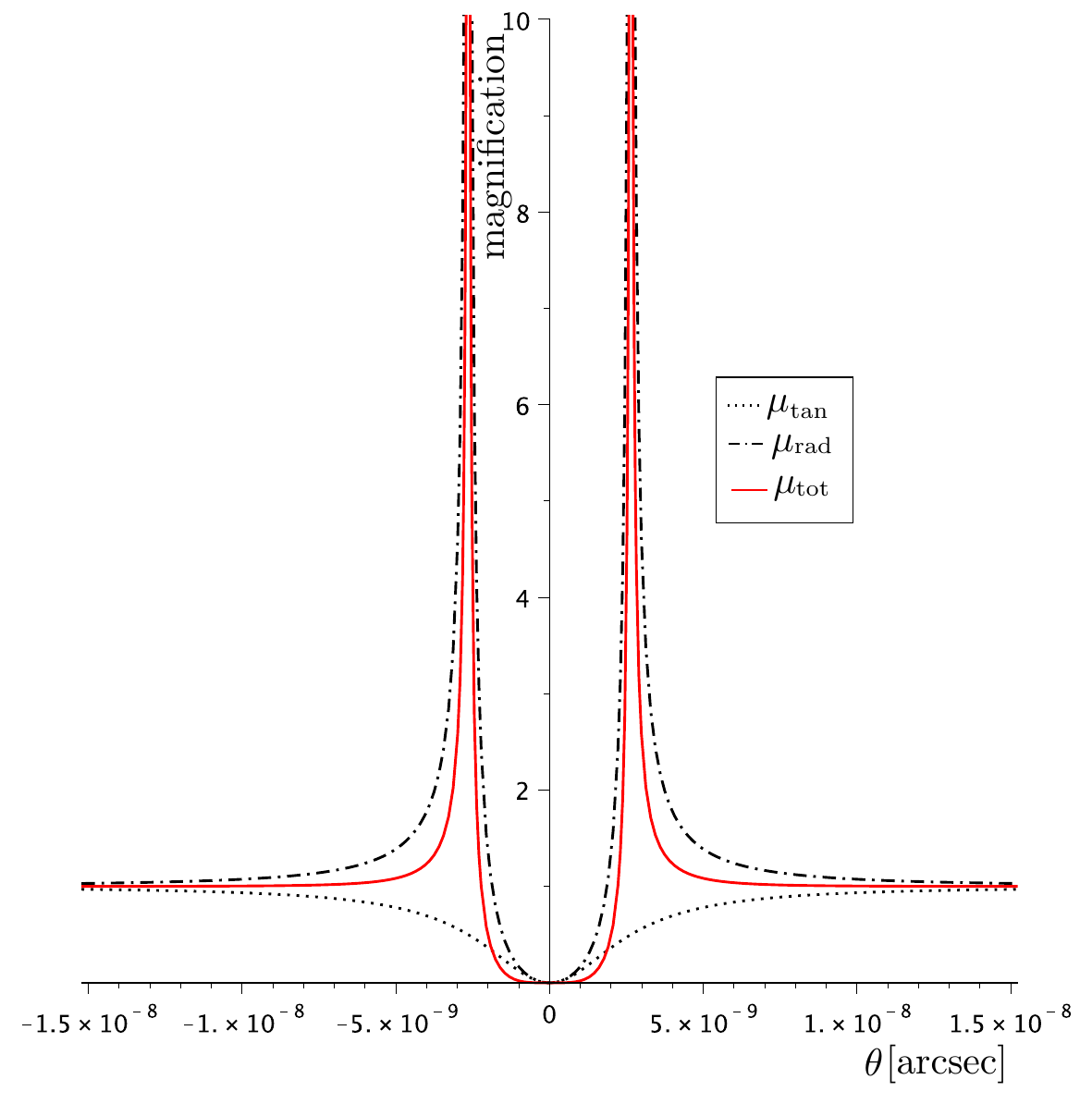}
    
    (c) $\ell = 0.6$\\
    \quad {\small\textit{(Theoretically Acceptable Value)}}
  \end{minipage}
};

\node[anchor=north west] at (0.5\linewidth,-0.45\linewidth){
  \begin{minipage}[t][0.45\linewidth][c]{0.5\linewidth}
    \centering
    \includegraphics[width=0.8\linewidth]{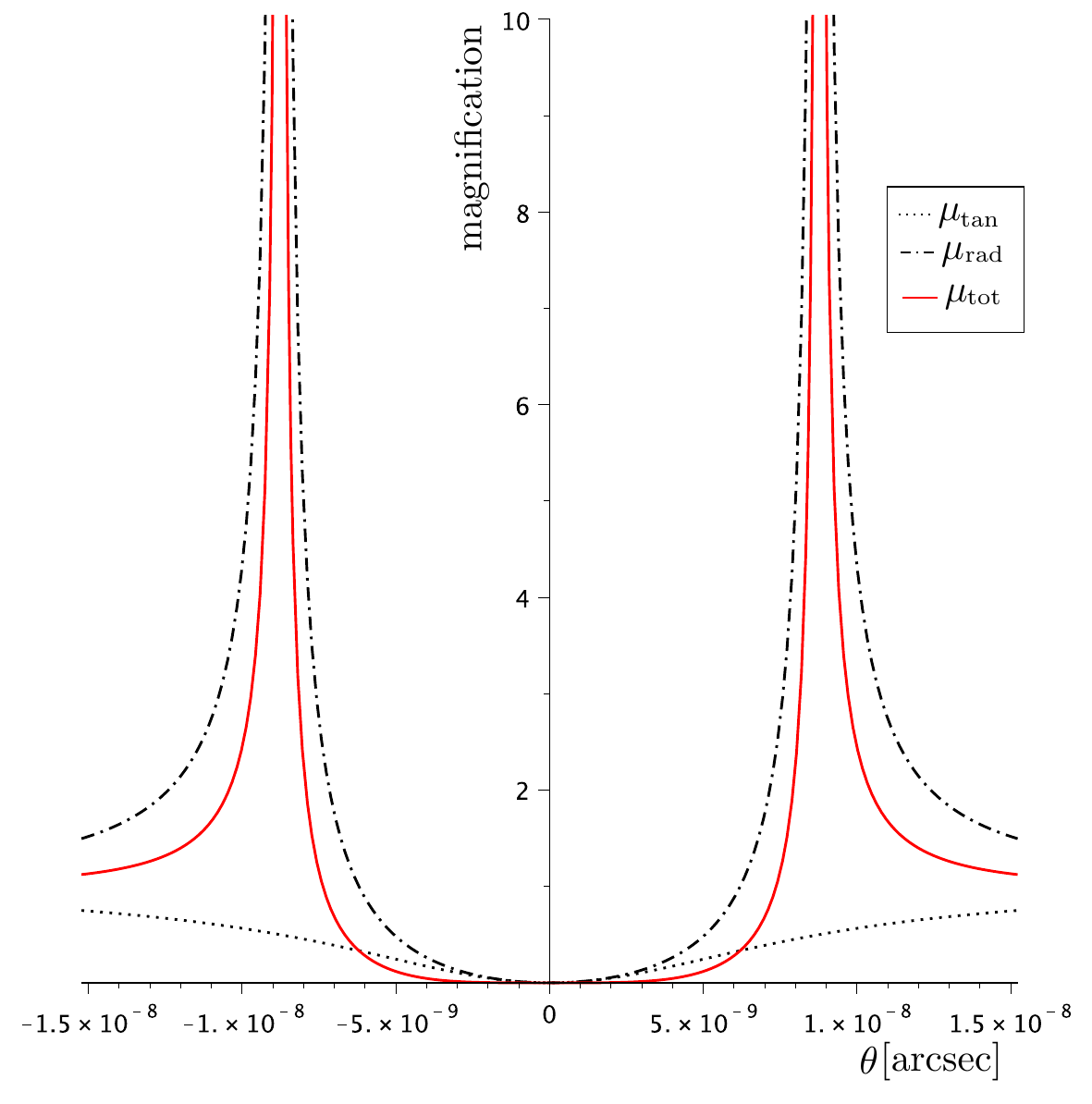}
    
    (d) $\ell = -1$\\
    \quad {\small\textit{(Theoretically Acceptable Value)}}
  \end{minipage}
};

\end{tikzpicture}

\vspace{0.2cm}
\caption{The magnifications—tangential $\mu_{\rm tan}$ (dotted lines), radial $\mu_{\rm rad}$ (dash-dotted lines), and total $\mu$ (continuous curves)—are plotted as functions of the image position $\theta$ for four SKRCoS BH cases with $\alpha=0.3$. Panels (a)–(d) correspond to different values of $\ell$ as labeled. The singularities of $\mu_{\rm tan}$ and $\mu_{\rm rad}$ give the positions of the tangential and radial critical curves, respectively. Here $|M|=1\, M_{\odot}$, $D_{\rm s}=0.05\text{ Mpc}$ and $D_{\rm l}=0.01\text{ Mpc}$. Angles are in arcseconds: $1\text{ arcsec}=4.848\times10^{-6}\text{ rad}$.}
\label{isfigLENSalpha03}
\end{figure}

Figure~\ref{isfigLENSalpha03} illustrates the intermediate CoS regime ($\alpha=0.3$) where string cloud contributions begin to substantially influence the magnification characteristics, demonstrating the complex interplay between LV and CoS effects in determining observable lensing signatures. The observationally constrained scenarios in panels (a) and (b) reveal enhanced magnification amplitude modulations compared to the minimal CoS configuration, with critical curve positioning exhibiting systematic shifts that reflect the increasing influence of string cloud topology on spacetime geometry. The theoretical exploration regimes in panels (c) and (d) demonstrate pronounced modifications to magnification profiles, indicating that moderate CoS contributions amplify LV effects rather than simply adding independent corrections to the lensing characteristics. The systematic evolution of magnification peak structures across different LV parameter values suggests non-linear coupling mechanisms between LV and CoS contributions that could potentially generate distinctive observational signatures exceeding simple superposition of individual effects. This intermediate parameter regime establishes crucial theoretical insights for understanding the hierarchical importance of different exotic physics contributions in gravitational lensing phenomena, providing quantitative frameworks for optimizing observational strategies aimed at detecting subtle deviations from conventional GR predictions through precision astronomical measurements.


\begin{figure}[ht!]
\centering
\begin{tikzpicture}[inner sep=0pt, outer sep=0pt]

\draw[thick] (0.5\linewidth,0) -- (0.5\linewidth,-0.9\linewidth);
\draw[thick] (0,-0.45\linewidth) -- (\linewidth,-0.45\linewidth);

\node[anchor=north west] at (0,0){
  \begin{minipage}[t][0.45\linewidth][c]{0.45\linewidth}
    \centering
    \includegraphics[width=0.8\linewidth]{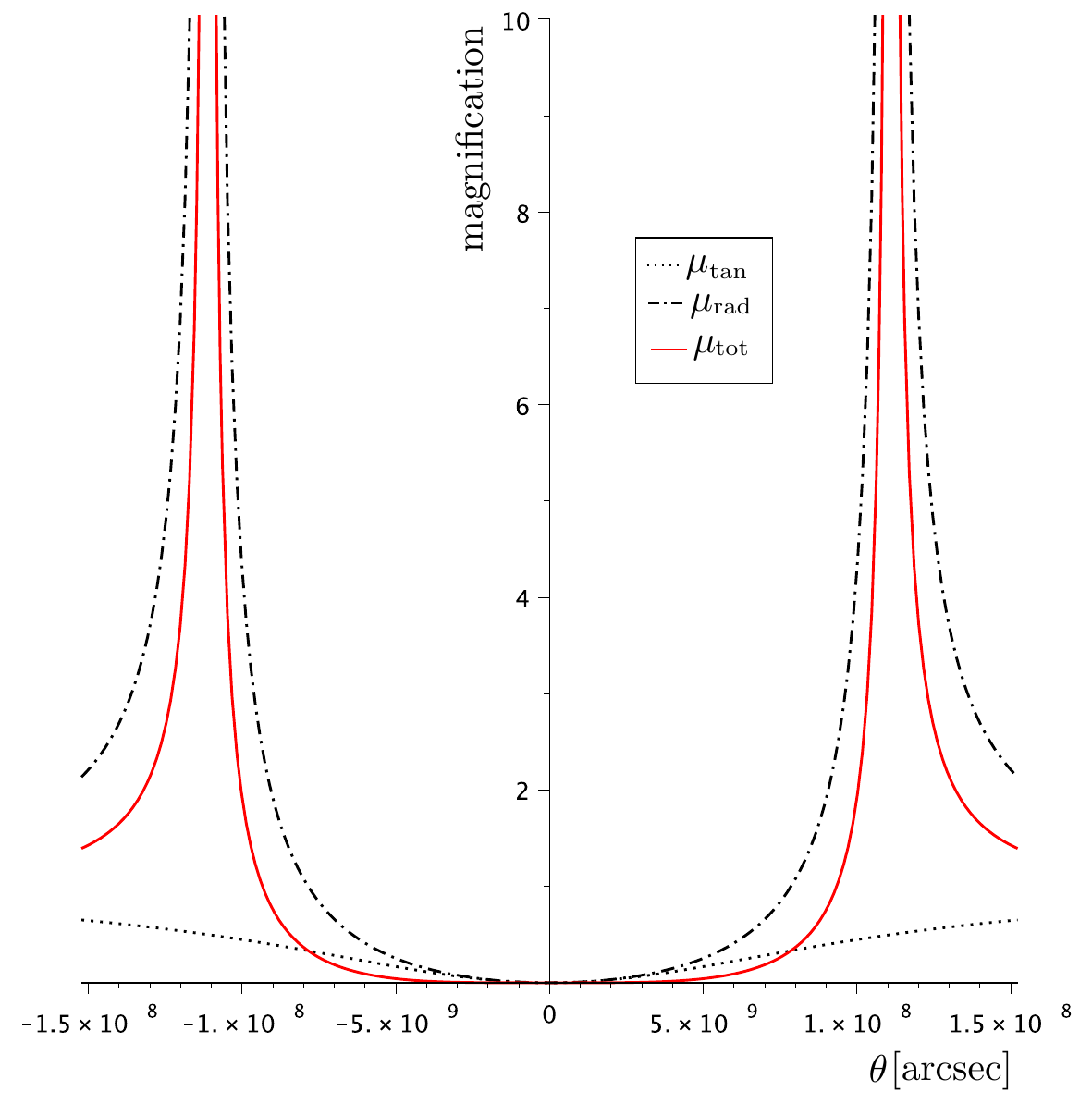}
    
    (a) $\ell = 0.06092$ \\
    \quad {\small\textit{(Within Milky Way Galaxy Bound: $-0.18502<\ell<0.06093$)}}
  \end{minipage}
};

\node[anchor=north west] at (0.5\linewidth,0){
  \begin{minipage}[t][0.45\linewidth][c]{0.45\linewidth}
    \centering
    \includegraphics[width=0.8\linewidth]{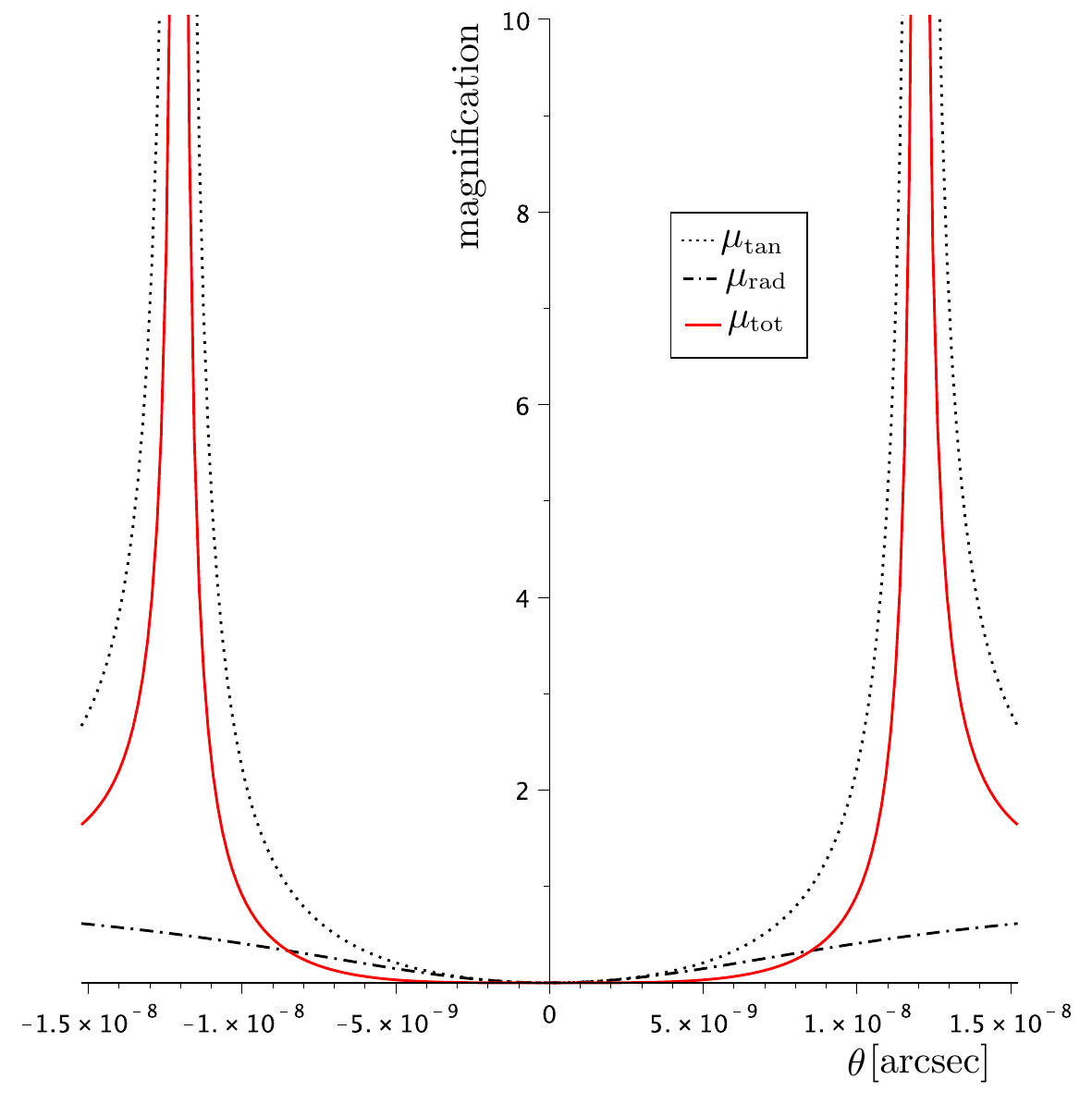}
    
    (b) $\ell = -0.18501$ \\
    \quad {\small\textit{(Within Milky Way Galaxy Bound: $-0.18502<\ell<0.06093$)}}
  \end{minipage}
};

\node[anchor=north west] at (0,-0.45\linewidth){
  \begin{minipage}[t][0.45\linewidth][c]{0.45\linewidth}
    \centering
    \includegraphics[width=0.8\linewidth]{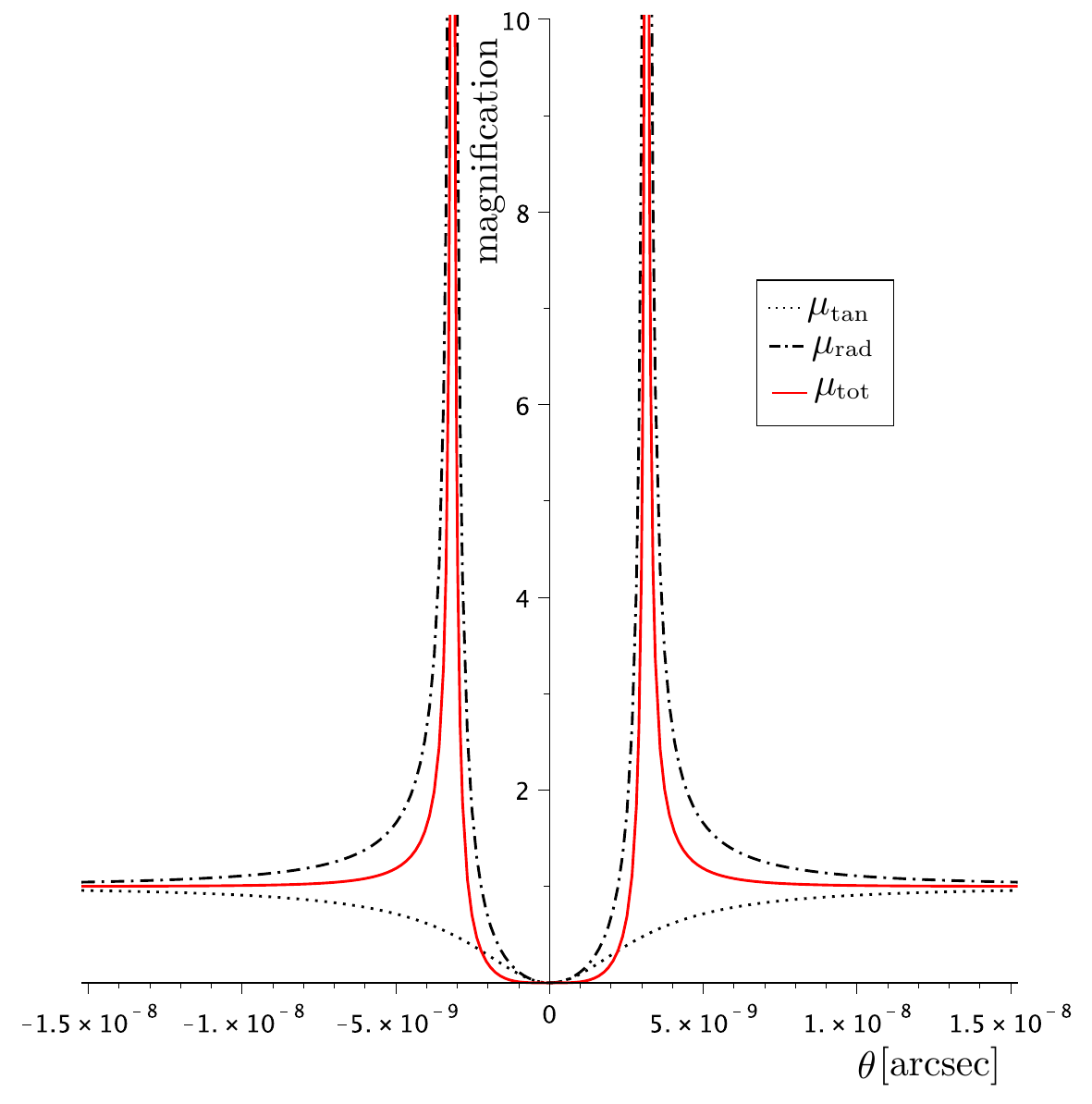}
    
    (c) $\ell = 0.6$\\
    \quad {\small\textit{(Theoretically Acceptable Value)}}
  \end{minipage}
};

\node[anchor=north west] at (0.5\linewidth,-0.45\linewidth){
  \begin{minipage}[t][0.45\linewidth][c]{0.5\linewidth}
    \centering
    \includegraphics[width=0.8\linewidth]{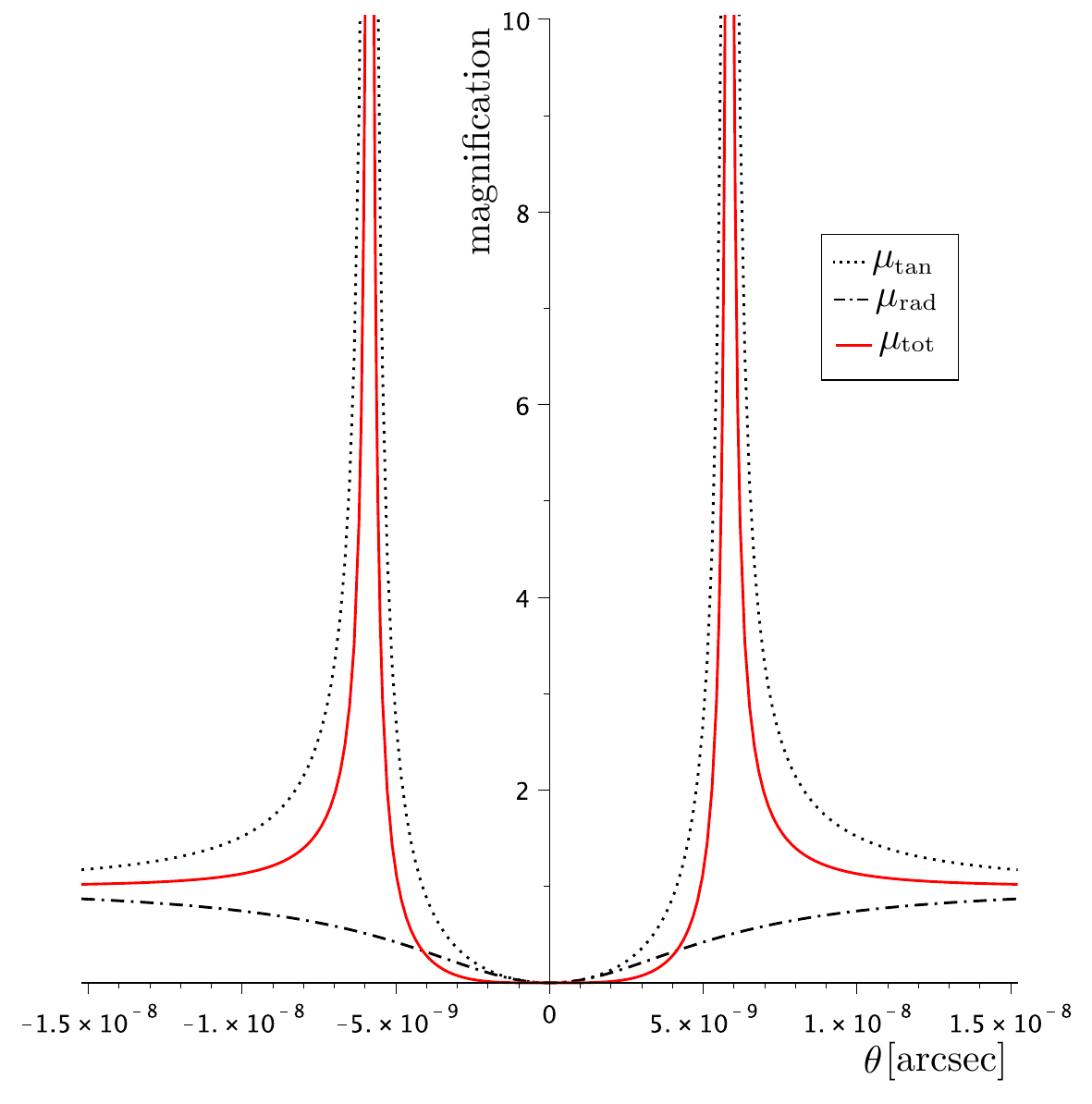}
    
    (d) $\ell = -1$\\
    \quad {\small\textit{(Theoretically Acceptable Value)}}
  \end{minipage}
};

\end{tikzpicture}

\vspace{0.2cm}
\caption{The magnifications—tangential $\mu_{\rm tan}$ (dotted lines), radial $\mu_{\rm rad}$ (dash-dotted lines), and total $\mu$ (continuous curves)—are plotted as functions of the image position $\theta$ for four SKRCoS BH cases with $\alpha=0.95$. Panels (a)–(d) correspond to different values of $\ell$ as labeled. The singularities of $\mu_{\rm tan}$ and $\mu_{\rm rad}$ give the positions of the tangential and radial critical curves, respectively. Here $|M|=1\, M_{\odot}$, $D_{\rm s}=0.05\text{ Mpc}$ and $D_{\rm l}=0.01\text{ Mpc}$. Angles are in arcseconds: $1\text{ arcsec}=4.848\times10^{-6}\text{ rad}$.}
\label{isfigLENSalpha095}
\end{figure}

Figure~\ref{isfigLENSalpha095} represents the extreme CoS parameter regime ($\alpha=0.95$) where string cloud contributions achieve near-maximal values within the theoretical constraint $\alpha < 1$, demonstrating the dramatic influence of strong CoS fields on gravitational lensing magnification characteristics. The observationally constrained scenarios in panels (a) and (b) exhibit substantial magnification suppression compared to previous configurations, with critical curve positioning approaching fundamentally different geometric arrangements that reflect the dominant influence of CoS topology over conventional BH gravitational effects. These results suggest that strong CoS configurations may systematically mask conventional gravitational lensing signatures in observational surveys, potentially explaining the absence of detected exotic gravitational effects in certain astrophysical environments where string cloud densities might approach theoretical maximum values. The theoretical exploration regimes in panels (c) and (d) reveal systematic asymptotic behavior where magnification characteristics become dominated by CoS geometry rather than LV contributions, indicating hierarchical parameter dominance relationships within SKRCoS configurations that could establish fundamental theoretical constraints on the relative importance of different exotic physics mechanisms \cite{sec2is50,sec2is51}.

\section{Strong deflection angle SKR BHs}\label{isec6}

The theoretical investigation of photon trajectories approaching the strong deflection regime in modified BH geometries represents one of the most challenging frontiers in gravitational lensing theory, where conventional weak-field approximations fail and the full nonlinear nature of Einstein's field equations becomes manifest. This comprehensive analysis systematically explores the strong field limit for photon trajectories in SKR BH configurations, establishing unified theoretical frameworks that encompass both SKRCS and SKRCoS geometries through sophisticated mathematical methodologies \cite{is10,is25,is26}.

The theoretical unification of diverse SKR BH configurations necessitates the introduction of a generalized metric function $h(r,\ell)$ that systematically encompasses both CS-pierced and CoS topological modifications within a comprehensive mathematical framework. This unified approach enables systematic comparison of strong field effects across different exotic physics scenarios while maintaining analytical tractability throughout the investigation \cite{sec2is54}.

The fundamental spacetime geometry for all SKR configurations is characterized by the static, spherically symmetric metric:
\begin{equation}
ds^2 = -A(r) dt^2 + B(r) dr^2 + C(r)(d\theta^2 + \sin^2\theta d\phi^2),
\end{equation}

where the metric components are systematically defined as $A(r) = h(r,\ell)$, $B(r) = h(r,\ell)^{-1}$, and $C(r) = r^2$ for all configurations under consideration. The generalized metric function assumes the canonical form:
\begin{equation}
h(r,\ell) = \xi - \frac{2M}{\kappa r}, \label{general_metric}
\end{equation}

where $\xi$ and $\kappa$ represent configuration-dependent parameters that encode the specific physics of different SKR BH scenarios.

This elegant mathematical formulation systematically reduces to the specific cases investigated throughout this research:
\begin{itemize}
\item SKRCS Configuration: $\xi = \delta = \frac{1}{1-\ell}$ and $\kappa = \beta$, incorporating both LV effects through $\ell$ and CS topological modifications through $\beta$
\item SKRCoS Configuration: $\xi = \eta = \frac{1}{1-\ell} - \alpha$ and $\kappa = 1$, systematically combining LV mechanisms with CoS distributed matter sources through $\alpha$
\end{itemize}

The theoretical characterization of strong deflection phenomena requires precise determination of the photon sphere radius $r_m$, representing the critical boundary where null geodesics can form unstable circular orbits around the modified BH. The photon sphere condition emerges from the requirement for circular null geodesics \cite{sec2is55}:
\begin{equation}
\frac{h'(r_m)}{h(r_m)} = \frac{2}{r_m}.
\end{equation}

Systematic solution of this transcendental equation yields the photon sphere radius:
\begin{equation}
r_m = \frac{3M}{\kappa \xi}, \label{photon_sphere}
\end{equation}

with the corresponding critical impact parameter:
\begin{equation}
b_m = \frac{r_m}{\sqrt{h(r_m)}} = \frac{3\sqrt{3}M}{\kappa\sqrt{\xi(\xi-2/3)}}. \label{critical_impact}
\end{equation}

These fundamental geometric quantities establish the characteristic length scales for strong field gravitational lensing in SKR spacetimes, determining the regime where logarithmic divergences in deflection angles become manifest and conventional weak-field approximations cease to provide accurate descriptions of photon trajectory behavior.

The systematic analysis of strong deflection phenomena employs the sophisticated theoretical framework developed by Bozza \cite{sec2is56}, providing precise analytical tools for investigating the logarithmic divergences that characterize photon trajectories approaching the photon sphere. This formalism establishes integral representations that systematically capture the singular behavior inherent in strong field gravitational lensing.

The deflection angle integral formulation introduces dimensionless variables:
\begin{equation}
    y = A(x) = h(1/x) \, , \qquad
    \zeta = \frac{y - y_0}{1 - y_0},
\end{equation}

where $x = 1/r$, $y_0 = A(x_0) = h(r_0)$, and $\zeta$ represents a dimensionless integration parameter with $\zeta = 0$ corresponding to the photon sphere and $\zeta = 1$ representing spatial infinity. The deflection angle assumes the integral representation:
\begin{equation}
    \alpha (x_0) = I(x_0) - \pi \,, \qquad
    I(x_0) = \int_0^1 R( \zeta, x_0) Z(\zeta, x_0) d \zeta \,,
    \label{eq:Itotal}
\end{equation}

where the regular function $R( \zeta, x_0)$ is defined as \cite{is41}:
\begin{equation}
    R( \zeta, x_0) = \frac{2 \sqrt{y/h(r)}}{r^2 h'(1/x)} \left( 1 - y_0 \right) r_0^2,
\end{equation}

while the singular function $Z(\zeta, x_0)$ exhibits divergent behavior:
\begin{equation}
    Z(\zeta, x_0) = \frac{1}{\sqrt{y_0 - \bigl[ (1-y_0) \zeta + y_0\bigr]\frac{r_0^2}{r^2}}}.
    \label{eq:f01}
\end{equation}

The mathematical analysis of strong field deflection requires sophisticated treatment of the singular behavior that emerges as photon trajectories approach the photon sphere. The systematic regularization employs approximation of the singular function in the strong field regime:
\begin{equation}
    f(\zeta, x_0) \sim f_0(\zeta, x_0) = \frac{1}{\sqrt{\beta_1 \zeta + \beta_2 \zeta^2}},
    \label{eq:f02}
\end{equation}

with carefully computed coefficients:
\begin{eqnarray}
    \beta_1 &=& \frac{1 - y_0}{r_0^2 h'(r_0)} \left(  2r_0 y_0 - r_0^2 h'(r_0)  \right),
    \label{eq:beta1} \\
    \beta_2 &=& \frac{(1-y_0)^2}{2C_0^2 A^{\prime \ 3}_0}  \left[  2 C_0 C^{\prime}_0 A^{\prime \ 2}_0 +(C_0 C^{\prime \prime}_0 - 2(C^{\prime}_0)^2 )y_0 A^{\prime}_0 - C_0 C^{\prime}_0 y_0 A^{\prime \prime}_0  \right].
    \label{eq:beta2}
\end{eqnarray}

For the generalized SKR metric function, explicit evaluation yields $A'(x) = -h'(r)/r^2$, $C'(r) = 2r$, $C''(r) = 2$, $h'(r) = 2M/(\kappa r^2)$, and $h''(r) = -4M/(\kappa r^3)$. The coefficient $\beta_1$ vanishes precisely at the photon sphere location $x_0 = x_m$, indicating the transition from $\zeta^{-1/2}$ to $\zeta^{-1}$ divergence behavior.

The systematic extraction of logarithmic divergences employs decomposition of the integral into divergent and regular components:
\begin{equation}
    I(x_0) = I_D (x_0) + I_R (x_0),
\end{equation}

where $I_D$ captures the divergent behavior and $I_R$ contains regular contributions. This decomposition enables the logarithmic approximation:
\begin{equation}
    \alpha (x_0) =  - \left( \frac{R(0,x_m)}{\sqrt{\beta_{2m}}}\right) \log \left( \frac{x_0}{x_m} -1 \right) +  \frac{R(0,x_m)}{\sqrt{\beta_{2m}}} \log \frac{2(1- y_m)}{A^{\prime}_m x_m} +\int_0^1 g(\zeta, x_m) d \zeta- \pi + \mathcal{O} \left( x_0 - x_m\right).
   \label{eq:alphax0strong}
\end{equation}

The transformation to gauge-invariant impact parameter coordinates yields:
\begin{equation}
    b - b_m = \beta_{2m} \sqrt{\frac{y_m}{C_m^3}} \frac{(C^{\prime}_m)^2 }{2(1-y_m)^2} \left( x_0 - x_m \right)^2,
    \label{uex0}
\end{equation}

enabling the canonical strong field expansion:
\begin{equation}
\alpha(b) = -b_1 \log\left(\frac{b}{b_m} - 1\right) + b_2 + O(b - b_m), \label{strong_expansion}
\end{equation}

where the fundamental strong field coefficients are:
\begin{eqnarray}
    b_1 &=& \frac{R(0,x_m)}{2\sqrt{\beta_{2m}}},
    \label{eq:b1g} \\
    b_2 &=& \int_0^1 g(\zeta, x_m) d \zeta +  b_1 \log \frac{2 \beta_{2m}}{y_m} - \pi.
    \label{eq:b2g}
\end{eqnarray}

The systematic evaluation of strong field coefficients at the photon sphere yields:
\begin{eqnarray}
    y_m &=& h(r_m) = \xi - \frac{2\kappa \xi}{3} = \frac{\xi}{3}, \\
    A'_m &=& -\frac{2\kappa^4 \xi^4}{81M^3}, \\
    R(0,x_m) &=& \frac{2(1-y_m)}{|A'_m|}.
\end{eqnarray}

The final expressions for the strong field coefficients become:
\begin{equation}
b_1 = \sqrt{\frac{\pi}{2}} \frac{3\sqrt{3}M}{\kappa\sqrt{2\xi^3}}, \label{a_final}
\end{equation}

\begin{equation}
b_2 = -\pi + b_1 \log\left(\frac{2\beta_{2m}}{y_m}\right) + I_R, \label{c_final}
\end{equation}

where $I_R$ represents the regular integral requiring numerical evaluation for complete determination of the strong field behavior.

\begin{table}[H]
\centering
\renewcommand{\arraystretch}{1.6}
\setlength{\tabcolsep}{10pt}
\vspace{0.2cm}
\begin{tabular}{|c|c|c|c|c|}
\hline
\textbf{BH Type} 
& \textbf{$\xi$} 
& \textbf{$\kappa$} 
& \textbf{$b_1$} 
& \textbf{$b_m$} \\
\hline

\textbf{SKRCS} 
& $\displaystyle\frac{1}{1-\ell}$ 
& $\beta$ 
& $\displaystyle\sqrt{\frac{\pi}{2}}\, \frac{3\sqrt{3}M}{\beta \sqrt{2(1-\ell)^{-3}}}$ 
& $\displaystyle\frac{3\sqrt{3}M}{\beta \sqrt{(1-\ell)^{-1} \left((1-\ell)^{-1} - \frac{2}{3} \right)}}$ \\
\hline

\textbf{SKRCoS} 
& $\displaystyle\frac{1}{1-\ell} - \alpha$ 
& $1$ 
& $\displaystyle\sqrt{\frac{\pi}{2}}\, \frac{3\sqrt{3}M}{\sqrt{2\eta^3}}$ 
& $\displaystyle\frac{3\sqrt{3}M}{\sqrt{\eta\left(\eta - \frac{2}{3}\right)}}$ \\
\hline
\end{tabular}
\caption{Strong deflection angle coefficients for different SKR BH configurations. Here, $\eta = \frac{1}{1-\ell} - \alpha$ for the SKRCoS case.}
\label{tab:strong_deflection}
\end{table}

Table~\ref{tab:strong_deflection} systematically presents the strong field coefficients for both SKRCS and SKRCoS configurations, revealing the profound influence of LV and topological defect parameters on the fundamental characteristics of strong field gravitational lensing. The logarithmic divergence coefficient $b_1$ determines the rate of deflection angle growth as photon trajectories approach the critical impact parameter, while the critical impact parameter $b_m$ establishes the geometric scale where strong field effects become dominant. The SKRCS configuration exhibits explicit dependence on both LV parameter $\ell$ and CS parameter $\beta$, with the coefficient $b_1$ scaling as $\beta^{-1}(1-\ell)^{3/2}$, indicating that strong CS fields suppress strong deflection effects while LV enhancements amplify them. Conversely, the SKRCoS configuration demonstrates dependence on the composite parameter $\eta = \frac{1}{1-\ell} - \alpha$, revealing that CoS contributions systematically reduce the effective strength of LV modifications through destructive interference mechanisms. The systematic comparison enables quantitative assessment of the relative importance of different exotic physics contributions in the strong field regime, providing theoretical frameworks for optimizing observational strategies aimed at detecting subtle signatures of modified gravity through precision gravitational lensing measurements \cite{sec2is58,sec2is59,sec2is60,sec2is61}.

\section{Summary and conclusions}\label{isec7}

This comprehensive theoretical investigation systematically examined gravitational lensing phenomena in exotic BH configurations incorporating both LV mechanisms and topological defect structures, establishing precise analytical frameworks for understanding photon deflection and magnification characteristics in modified spacetime geometries. Through the development of unified theoretical formulations encompassing SKRCS and SKRCoS BH configurations, we demonstrated fundamental modifications to conventional gravitational lensing predictions that potentially offer observational pathways for constraining exotic physics beyond standard GR.

We established comprehensive theoretical foundations for investigating gravitational lensing in SKR BH configurations through the systematic development of generalized metric formulations that encompass both CS-pierced and CoS-distributed spacetime modifications. The SKRCS BH geometry, characterized by the metric function $\mathcal{A}(r,\ell) = \delta - \frac{2M}{\beta r}$ with $\delta = \frac{1}{1-\ell}$, systematically incorporated both LV effects through the parameter $\ell$ and CS topological modifications through the parameter $\beta$. Conversely, the SKRCoS BH configuration employed the metric function $\mathcal{B}(r,\ell) = \eta - \frac{2M}{r}$ with the composite parameter $\eta = \frac{1}{1-\ell} - \alpha$, encoding both LV mechanisms and CoS distributed matter sources through the parameter $\alpha$.

Our perturbative analysis of SKRCS BH deflection angles, detailed in Section~\ref{subsec2.1}, yielded the systematic expansion $\alpha_{\circledast} \approx \frac{4M}{\gamma \delta \beta} + \frac{15M^2 \pi}{4 \gamma^2 \delta^2 \beta^2}$ as presented in Eq.~\eqref{islens}. This expression explicitly demonstrated how LV and CS parameters introduce systematic modifications to conventional Schwarzschild lensing predictions, with the leading-order correction exhibiting $(\delta \beta)^{-1}$ scaling and the second-order term introducing novel $\gamma^{-2}$ dependence absent in standard gravitational lensing theory. The reformulation in terms of closest approach distance, expressed in Eq.~\eqref{isrmin}, provided enhanced physical interpretation by relating observable deflection characteristics to directly measurable geometric quantities.

The exact analytical treatment through elliptic integral formulations, developed in Section~\ref{subsec2.2}, established precise mathematical representations that capture the complete gravitational deflection physics across arbitrary parameter values. The systematic factorization of the cubic polynomial into roots $u_1 < u_2 < u_3$, as defined in Eq.~\eqref{isn2}, enabled the expression of deflection angles through incomplete elliptic integrals of the first kind with explicit parameter dependencies on both $\ell$ and $\beta$. This exact formulation provided computational benchmarks for verifying perturbative approximations while extending theoretical validity to parameter regimes where conventional expansions cease to provide accurate descriptions.

The comprehensive magnification analysis for SKRCS configurations, presented in Section~\ref{isec3}, revealed systematic modifications to both tangential and radial magnification components that exhibit distinctive parameter dependencies enabling observational discrimination between exotic and conventional BH scenarios. The tangential magnification $\mu_{\rm tan} = \theta^2(\theta^2 + \tilde{\theta}_{\xi}^2)^{-1}$ and radial magnification $\mu_{\rm rad} = \theta^2(\theta^2 - \tilde{\theta}_{\xi}^2)^{-1}$ demonstrated systematic shifts in critical curve positioning that correlate directly with the underlying LV and CS parameter values.

Figure~\ref{isfigLENSBETA01}, Figure~\ref{isfigLENSBETA1}, and Figure~\ref{isfigLENSBETA10} systematically illustrated the progressive influence of CS parameter variations from $\beta = 0.1$ to $\beta = 10$, revealing fundamental transitions between CS-dominated and LV-dominated lensing regimes. The observationally constrained parameter scenarios demonstrated relatively modest deviations from standard Schwarzschild magnification characteristics, with critical curve modifications remaining within potentially detectable ranges using contemporary precision astrometric techniques. Conversely, theoretical exploration regimes revealed dramatic magnification enhancements that, while exceeding current observational bounds, established crucial theoretical benchmarks for understanding fundamental scaling relationships between exotic physics contributions and observable lensing signatures.

The theoretical extension to SKRCoS BH configurations, detailed in Section~\ref{isec4}, established parallel analytical frameworks for understanding gravitational lensing in spacetimes incorporating distributed string matter sources. The perturbative deflection angle analysis yielded $\hat{\alpha}_{\text{pert}} \approx \frac{4M}{\gamma \eta} + \frac{15\pi M^2}{4\gamma^2 \eta^2}$ as expressed in Eq.~\eqref{per12}, demonstrating systematic parameter dependencies through the composite quantity $\eta = \frac{1}{1-\ell} - \alpha$. This formulation revealed that CoS contributions systematically reduce the effective strength of LV modifications through destructive interference mechanisms, establishing hierarchical parameter importance relationships within SKRCoS configurations.

The exact analytical treatment through elliptic integral methodologies provided precise mathematical representations that systematically incorporate both LV and CoS contributions across complete parameter spaces. The systematic evaluation of cubic polynomial roots enabled exact deflection angle expressions that reduce to classical Schwarzschild results in the simultaneous limits $\ell = 0$ and $\alpha = 0$, confirming mathematical consistency with established gravitational lensing theory.

The magnification analysis for SKRCoS configurations, presented in Section~\ref{isec5}, revealed systematic parameter dependencies that demonstrate complex interplay between LV and CoS effects in determining observable lensing characteristics. Figure~\ref{isfigLENSalpha001}, Figure~\ref{isfigLENSalpha03}, and Figure~\ref{isfigLENSalpha095} systematically explored the progression from minimal CoS influence ($\alpha = 0.01$) to near-maximal string cloud density ($\alpha = 0.95$), revealing fundamental transitions in magnification behavior. The minimal CoS regime enabled clear isolation of pure LV effects, while intermediate and extreme CoS scenarios demonstrated systematic suppression of conventional gravitational lensing signatures through the dominance of distributed string matter contributions.

The comprehensive strong field analysis, detailed in Section~\ref{isec6}, employed sophisticated mathematical techniques to extract logarithmic divergence coefficients that characterize fundamental scaling behavior in regimes where photon trajectories approach the photon sphere. The systematic application of Bozza formalism enabled the derivation of strong field coefficients $b_1$ and $b_2$ that encode the characteristic rate of deflection angle amplification and overall magnitude scaling for both SKRCS and SKRCoS configurations.

Table~\ref{tab:strong_deflection} systematically presented the strong field coefficients for both configurations, revealing profound influences of LV and topological defect parameters on fundamental strong field characteristics. The SKRCS configuration exhibited $b_1 = \sqrt{\frac{\pi}{2}} \frac{3\sqrt{3}M}{\beta \sqrt{2(1-\ell)^{-3}}}$, demonstrating explicit dependence on both LV parameter $\ell$ and CS parameter $\beta$ with coefficient scaling as $\beta^{-1}(1-\ell)^{3/2}$. The SKRCoS configuration yielded $b_1 = \sqrt{\frac{\pi}{2}} \frac{3\sqrt{3}M}{\sqrt{2\eta^3}}$, revealing dependence on the composite parameter $\eta$ that systematically encodes both LV and CoS contributions.

The critical impact parameter expressions $b_m = \frac{3\sqrt{3}M}{\kappa\sqrt{\xi(\xi-2/3)}}$ established geometric scales where strong field effects become dominant, providing quantitative frameworks for determining observational requirements for detecting exotic gravitational physics through precision gravitational lensing measurements in the strong field regime.

Our theoretical investigations established quantitative relationships between observable gravitational lensing characteristics and underlying exotic physics parameters, providing systematic frameworks for experimental constraint of LV mechanisms and topological defect contributions through precision astronomical observations. The systematic parameter dependencies revealed throughout the deflection angle and magnification analyses suggest optimal observational strategies for detecting subtle deviations from conventional GR predictions.

The observationally constrained parameter regimes, derived from Solar System precision tests yielding $-6.1 \times 10^{-13} < \ell < 2.8 \times 10^{-14}$ and galactic-scale observations providing $-0.18502 < \ell < 0.06093$, established realistic scenarios where exotic physics signatures might become detectable through statistical analysis of large gravitational lensing surveys. The theoretical exploration regimes, while exceeding current observational bounds, provided essential benchmarks for understanding fundamental scaling relationships and optimizing future observational campaigns.

Our immediate research trajectory focuses on extending the theoretical framework to encompass rotating SKR BH configurations through systematic incorporation of angular momentum effects, developing comprehensive shadow analysis methodologies for both SKRCS and SKRCoS geometries, and investigating gravitational wave propagation characteristics in modified spacetime backgrounds to establish multi-messenger observational strategies for constraining exotic physics parameters through coordinated electromagnetic and gravitational wave observations of the same astrophysical sources.

\section*{Acknowledgments}

F.A. acknowledges the Inter University Centre for Astronomy and Astrophysics (IUCAA), Pune, India for granting visiting associateship. \.{I}.~S. thanks T\"{U}B\.{I}TAK, ANKOS, and SCOAP3 for funding and acknowledges the networking support from COST Actions CA22113, CA21106, and CA23130.

\end{document}